\documentclass[letterpaper,english,aps,letter,superscriptaddress,nofootinbib,floatfix,pre,preprint]{revtex4}
\pdfoutput=1
\usepackage[T1]{fontenc}
\usepackage[latin1]{inputenc}
\usepackage{graphicx}
\usepackage{amssymb}

\makeatletter

\providecommand{\boldsymbol}[1]{\mbox{\boldmath $#1$}}

\providecommand{\tabularnewline}{\\}

\usepackage{ae,aecompl}

\usepackage{babel}
\makeatother
\begin{document}

\title{An Event-Driven Hybrid Molecular Dynamics and Direct Simulation Monte
Carlo Algorithm}

\author{Aleksandar Donev}

\affiliation{Lawrence Livermore National Laboratory, P.O.Box 808, Livermore, CA
94551-9900}

\author{Alejandro L. Garcia}

\affiliation{Department of Physics, San Jose State University, San Jose, California,
95192}

\author{Berni J. Alder}

\affiliation{Lawrence Livermore National Laboratory, P.O.Box 808, Livermore, CA
94551-9900}

\begin{abstract}
A novel Stochastic Event-Driven Molecular Dynamics (SEDMD) algorithm
is developed for the simulation of polymer chains suspended in a solvent.
The polymers are represented as chains of hard spheres tethered by
square wells and interact with the solvent particles with hard core
potentials. The algorithm uses Event-Driven Molecular Dynamics (EDMD)
for the simulation of the polymer chain and the interactions between
the chain beads and the surrounding solvent particles. The interactions
between the solvent particles themselves are not treated deterministically
as in event-driven algorithms, rather, the momentum and energy exchange
in the solvent is determined stochastically using the Direct Simulation
Monte Carlo (DSMC) method. The coupling between the solvent and the
solute is consistently represented at the particle level, however,
unlike full MD simulations of both the solvent and the solute, the
spatial structure of the solvent is ignored. The algorithm is described
in detail and applied to the study of the dynamics of a polymer chain
tethered to a hard wall subjected to uniform shear. The algorithm
closely reproduces full MD simulations with two orders of magnitude
greater efficiency. Results do not confirm the existence of periodic
(cycling) motion of the polymer chain.
\end{abstract}
\maketitle
\newcommand{\Cross}[1]{\left|\mathbf{#1}\right|_{\times}}
\newcommand{\CrossL}[1]{\left|\mathbf{#1}\right|_{\times}^{L}}
\newcommand{\CrossR}[1]{\left|\mathbf{#1}\right|_{\times}^{R}}
\newcommand{\CrossS}[1]{\left|\mathbf{#1}\right|_{\boxtimes}}

\newcommand{\V}[1]{\mathbf{#1}}
\newcommand{\M}[1]{\mathbf{#1}}
\newcommand{\D}[1]{\Delta#1}

\newcommand{\sV}[1]{\boldsymbol{#1}}
\newcommand{\sM}[1]{\boldsymbol{#1}}

\newcommand{\grad}{\boldsymbol{\nabla}}
\newcommand{\eij}{\left\{  i,j\right\}  }

\newcommand{\Wi}{\mbox{Wi}}

\section{Introduction}

Driven by nanoscience interests it has become necessary to develop
tools for hydrodynamic calculations at the atomistic scale \cite{Nanohydrodynamics_Alder,DSMC_MPCD_Gompper,Microfluidics_Review,PolymerTumbling_Review,FluctuatingHydro_Coveney}.
Of particular interest is the modeling of polymers in a flowing {}``good''
solvent for both biological (e.g., cell membranes) and engineering
(e.g., micro-channel DNA arrays) applications \cite{PolymerDynamics_Review,PolymerTumbling_Review}.
The most widely studied polymer models are simple linear bead-spring;
freely-jointed rods; or worm-like chains. Such models have been parameterized
for important biological and synthetic polymers. Much theoretical,
computational, and experimental knowledge about the behavior of these
models has been accumulated for various representations of the solvent.
However, the multi-scale nature of the problem for both time and length
is still a challenge for simulation of reasonably large systems over
reasonably long times. Furthermore, the omission in these models of
the \emph{explicit} coupling between the solvent and the polymer chain(s)
requires the introduction of adjustable parameters (e.g., friction
coefficients) to be determined empirically. The algorithm presented
here overcomes this for a linear polymer chain tethered to a hard
wall and subjected to a simple linear shear flow \cite{TetheredPolymer_Experiment_PRL,TetheredPolymer_HybridMD,TetheredPolymer_FullMD,TetheredPolymer_Cyclic_PRL,TetheredDNA_FullMD}.
Of particular interest is the long-time dynamics of the polymer chain
\cite{TetheredPolymer_Experiment_PRL,TetheredPolymer_FullMD,PolymerTumbling_PRL,TetheredPolymer_Cyclic_PRL,TetheredPolymer_Cyclic_AIP}
and any effects of the polymer motion on the flow field.

Brownian dynamics is one of the standard methods for coupling the
polymer chains to the solvent \cite{BrownianDynamics_DNA,BrownianDynamics_OrderN}.
The solvent is only implicitly represented by a coupling between the
polymer beads and the solvent in the form of stochastic (white-noise)
forcing and linear frictional damping. The flow in the solvent is
not explicitly simulated, but approximated as a small perturbation
based on the Oseen tensor. This approximation is only accurate at
large separations of the beads and at sufficiently small Reynold's
numbers. Even algorithms that do model the solvent explicitly via
Lattice Boltzmann (LB) \cite{LatticeBoltzmann_Polymers}, incompressible
(low Reynolds number) CFD solvers \cite{FluctuatingHydro_FluidOnly,DNA_Laden_Flow,FluctuatingHydroMD_Coveney},
or multiparticle collision dynamics \cite{PolymerCollapse_Yeomans,Trebotich_HardRods,MultiparticleDSMC_Polymers,DSMC_MPCD_MD_Kapral},
typically involve phenomenological coupling between the polymer chain
and the flowing fluid in the form of a linear friction term based
on an effective viscosity (a notable exception being the algorithm
described in Ref. \cite{DSMC_MPCD_MD_Kapral}). Furthermore, solvent
fluctuations in the force on the polymer beads are often approximated
without fully accounting for spatial and temporal correlations. Finally,
the reverse coupling of the effect of the bead motion on the fluid
flow is either neglected or approximated with delta function forcing
terms in the continuum fluid solver \cite{Trebotich_Penalty}. More
fundamentally, continuum descriptions of flow at micro and nanoscales
are known to have important deficiencies \cite{Microfluidics_Review,Nanohydrodynamics_Alder}
and therefore it is important to develop an all-particle algorithm
that is able to reach the long times necessary for quantitative evaluation
of approximate, but faster, algorithms.

The most detailed (and expensive) modeling of polymers in flow is
explicit molecular dynamics (MD) simulation of both the polymer (solute)
and the surrounding solvent \cite{PolymerShear_MD,TetheredDNA_FullMD}.
Multi-scale algorithms have been developed to couple the MD simulation
to Navier-Stokes-based computational fluid dynamics (CFD) calculations
of the flow field \cite{TetheredPolymer_HybridMD}. However, the calculation
time still remains limited by the slow molecular dynamics component.
Thus the computational effort is wasted on simulating the structure
and dynamics of the solvent particles, even though it is the polymer
structure and dynamics (and their coupling to the fluid flow) that
is of interest. Our algorithm replaces the deterministic treatment
of the solvent-solvent interactions with a stochastic momentum exchange
operation, thus significantly lowering the computational cost of the
algorithm, while preserving microscopic details in the solvent-solute
coupling.

Fluctuations drive the polymer motion and must be accurately represented
in any model. Considerable effort has been invested in recent years
in including fluctuations directly into the Navier-Stokes (NS) equations
and the associated CFD solvers \cite{FluctuatingHydro_FluidOnly,FluctuatingHydro_Coveney,FluctuatingHydro_Garcia}.
Such fluctuating hydrodynamics has been coupled to molecular dynamics
simulations of polymer chains \cite{FluctuatingHydroMD_Coveney},
but with empirical coupling between the beads and the fluid as discussed
above. To avoid the empirical coupling, the solvent region could be
enlarged by embedding the atomistic simulations of the region around
the polymer chain (such as pure MD or our combined MD/DSMC algorithm)
in a fluctuating hydrodynamics region. The bidirectional coupling
between the continuum and particle regions has to be constructed with
great care so that both fluxes and fluctuations are preserved \cite{FluctuatingHydro_AMAR}.
A well-known problem with such multiscale approaches is that the finest
scale (atomistic simulation) can take up the majority of computational
time and thus slow down the whole simulation. By using DSMC the cost
of the particle region can be made comparable to that of the continuum
component.

The Stochastic Event-Driven Molecular Dynamics (SEDMD) algorithm presented
here combines Event-Driven Molecular Dynamics (EDMD) for the polymer
particles with Direct Simulation Monte Carlo (DSMC) for the solvent
particles. The polymers are represented as chains of hard spheres
tethered by square wells. The solvent particles are realistically
smaller than the beads and are considered as hard spheres that interact
with the polymer beads with the usual hard-core repulsion. The algorithm
processes true (deterministic, exact) binary collisions between the
solvent particles and the beads, without any approximate coupling
or stochastic forcing. However, for the purposes of the EDMD algorithm,
the solvent particles themselves do not directly interact with each
other, that is, they can freely pass through each other as for an
ideal gas. Deterministic collisions between the solvent particles
are replaced with stochastic DSMC collisions. Both asynchronous (event-driven)
and synchronous (time-driven) algorithmic ways of processing these
stochastic collisions will be discussed in the next section. Note
that our algorithm is similar to a recent algorithm developed for
soft interaction potentials combining time-driven MD with multiparticle
collision dynamics \cite{DSMC_MPCD_MD_Kapral}. 

The fundamental ideas behind our algorithm are described next, and
further details are given in Section \ref{sec:Algorithmic-Details}.
Section \ref{sec:Tethered-Polymer} gives results from the application
of the algorithm to the tethered polymer problem, and some concluding
remarks are given in Section \ref{sec:Conclusions}.

\section{Hybrid Components}

In this section we briefly describe the two components of the SEDMD
algorithm: The stochastic handling of the solvent and the deterministic
handling of the solute particles. These two components are integrated
(i.e., \emph{tightly} coupled) into a single event-driven algorithm
in Section \ref{sec:Algorithmic-Details}.

\subsection{Solvent DSMC Model}

The validity of the (incompressible) Navier-Stokes continuum equations
for modeling microscopic flows has been well established down to length
scales of $10-100nm$ \cite{Microfluidics_Review}. However, there
are several issues present in microscopic flows that are difficult
to account for in models relying on a purely PDE approximation. Firstly,
it is not \emph{a priori} obvious how to treat boundaries and interfaces
well so as account for the non-trivial (possibly non-linear) coupling
between the flow and the microgeometry. Furthermore, fluctuations
are not typically considered in Navier-Stokes solvers, and they can
be very important at instabilities \cite{FluidMixing_DSMC} or in
driving polymer dynamics. Finally, since the grid cell sizes needed
to resolve complex microscopic flows are small, a large computational
effort (comparable to DSMC) is needed even for continuum solvers.
An alternative is to use particle-based methods, which are explicit
and unconditionally stable and rather simple to implement. The solvent
particles are directly coupled to the microgeometry, for example,
they directly interact with the beads of a polymer. Fluctuations occur
naturally with the correct spatio-temporal correlations. However,
as in continuum descriptions, the structure of the fluid is lost and
under certain conditions the high compressibility of the DSMC (ideal
gas) fluid can cause difficulties.

Several particle methods have been described in the literature, such
as MD \cite{PolymerShear_MD}, dissipative particle dynamics (DPD)
\cite{DPD_DNA}, and multi-particle collision dynamics (MPCD) \cite{DSMC_MPCD_Gompper,DSMC_MPCD_MD_Kapral}.
Our method is similar to MPCD (also called stochastic rotation dynamics
or the Malevanets-Kapral method), and in fact, both are closely related
to the Direct Simulation Monte Carlo (DSMC) algorithm of Bird \cite{DSMCReview_Garcia}.
The key idea behind DSMC is to replace deterministic interactions
between the particles with stochastic momentum exchange (collisions)
between nearby particles. Specifically, particles are propagated by
a fixed time step $\D{t}$, as in MD, moving ballistically along straight
lines during a time-step (advection step). At the end of each time
step, the particles are sorted into cells, each containing on the
order of ten particles, and then a certain number of random pairs
of particles that are in the same cell are chosen to undergo stochastic
collisions (collision step). These collisions do not take into account
the positions of the particles other than the fact they are in the
same cell (i.e., they are nearby). The collisions conserve momentum
and energy (but not angular momentum) exactly. Formally, DSMC can
be seen as a method for solving the Boltzmann transport equation for
a low-density gas, however, it is not limited to gas flows \cite{DSMC_DenseFluids,DSMC_CBA,DSMC_CBATheory}.
Our purpose for using DSMC is as a replacement for expensive MD, preserving
the essential hydrodynamic {}``solvent'' properties: local momentum
conservation, and linear momentum exchange on length scales comparable
to the particle size, and a similar fluctuation spectrum.

In the multiparticle collision variant of this algorithm originally
proposed by Kapral, the traditional DSMC collection of binary collisions
is replaced by a multi-particle collision in which the velocities
of all particles in the cell are rotated by a random amount around
the average velocity \cite{DSMC_MPCD_Gompper,DSMC_MPCD_MD_Kapral}.
This change improves efficiency but at the cost of some artificial
effects such as loss of Galilean invariance. These problems can be
corrected and the method has been successfully used in modeling polymers
in flow by including the beads, considered as (massive) point particles,
in the stochastic momentum exchange step \cite{MultiparticleDSMC_Polymers,PolymerCollapse_Yeomans,ShearThinning_Yeomans}.
We will employ traditional DSMC in our algorithm in order to mimic
the actual (deterministic) momentum exchange between solvent molecules
(as it would be in an MD simulation) and in order to avoid any possible
artifacts.

A fundamental deficiency of DSMC as a (micro or nano) hydrodynamic
solver is the large (ideal gas) compressibility of the fluid. For
subsonic flows this compressibility does not qualitatively affect
the results as the DSMC fluid will behave similarly to an incompressible
liquid, however, the (Poisson) density fluctuations in DSMC are significantly
larger than those in realistic liquids. Furthermore, the speed of
sound is small (comparable to the average speed of the particles)
and thus subsonic (Mach number less than one) flows are limited to
relatively small Reynolds numbers %
\footnote{For a low-density gas the Reynolds number is $Re=M/K$, where $M=v_{flow}/c$
is the Mach number, and the Knudsen number $K=\lambda/L$ is the ratio
between the mean free path $\lambda$ and the typical obstacle length
$L$. This shows that subsonic flows can only achieve high $Re$ flows
for small Knudsen numbers, i.e., large numbers of DSMC particles.%
}. The Consistent Boltzmann Algorithm (CBA) \cite{DSMC_CBA,DSMC_CBATheory},
as well as algorithms based on the Enskog equation \cite{DSMC_Enskog,DSMC_Enskog_Frezzotti},
have demonstrated that DSMC fluids can have dense-fluid compressibility.
A similar algorithm was recently constructed for MPCD \cite{MPCD_CBA}.
We are currently evaluating several DSMC variants in terms of their
efficiency and thermodynamic consistency under high densities %
\footnote{Note that the density fluctuations in the CBA fluid are identical
to those in an ideal gas and thus thermodynamically inconsistent with
the compressibility.%
} and will report our findings in future work.

\subsection{Polymer MD Model}

Polymer chains in a solvent are modeled using continuous pair potentials
and time-driven MD (TDMD), in which particles are synchronously propagated
using a time step $\D{t}$, integrating the equations of motion along
the way. For good solvents, the polymer beads are represented as spherical
particles that interact with other beads and solvent particles with
(mostly) repulsive pair potentials, such as the positive part of the
Lennard-Jones potential. Additionally, beads are connected via (usually
finitely-extensible FENE or worm-like) springs in order to mimic chain
connectivity and elasticity \cite{PolymerShear_MD}. Additionally,
stochastic forces may be present to represent the solvent. The time
steps required for integration of the equations of motion in the presence
of the strongly repulsive forces is small and TDMD cannot reach long
time scales even after parallelization. An alternative is to use hard
spheres instead of soft particles, allowing replacement of the FENE
springs with square-well tethers, thus avoiding the costly force evaluations
in traditional MD. Hard sphere MD is most efficiently performed using
event-driven molecular dynamics (EDMD) \cite{EventDriven_Alder,Event_Driven_HE,EDMD_Polymers_Hall,PolymerCollapse_EDMD}.
If the detailed structure and energetics of the liquid is not crucial,
such EDMD algorithms can be just as effective as TDMD ones but considerably
faster. The essential difference between EDMD and TDMD is that EDMD
is asynchronous and there is no time step, instead, collisions between
hard particles are explicitly predicted and processed at their exact
(to numerical precision) time of occurrence. Since particles move
along simple trajectories (straight lines) between collisions, the
algorithm does not waste any time simulating motion in between events
(collisions).

Hard-sphere models of polymer chains have been used in EDMD simulations
for some time \cite{EDMD_Polymers_Hall,EDMD_Polymers_Aggregation,EDMD_Polymer_Fibrils2}.
These models typically involve, in addition to the usual hard-core
exclusion, additional \emph{square well} interactions to model chain
connectivity. The original work by Alder \emph{et al.} on EDMD developed
the collisional rules needed to handle arbitrary square wells \cite{EventDriven_Alder}.
Infinitely high wells can model tethers between beads, and the tethers
can be allowed to be broken by making the square wells of finite height,
modeling soft short-range attractions. Recent studies have used square
well attraction to model the effect of solvent quality \cite{PolymerCollapse_EDMD}.
Even more complex square well models have been developed for polymers
with chemical structure and it has been demonstrated that such models,
despite their apparent simplicity, can successfully reproduce the
complex packing structures found in polymer aggregation \cite{EDMD_Polymer_Fibrils2,EDMD_Polymers_Aggregation}.
Recent work on coupling a Kramer bead-rod polymer to a NS solver has
found that the use of hard rods (instead of soft interactions) not
only rigorously prevents rod-rod crossing but also achieves a larger
time step, comparable to the time step of the continuum solver \cite{Trebotich_HardRods}.

This study is focused on the simplest model of a polymer chain, namely,
a linear chain of $N_{b}$ particles tethered by unbreakable bonds.
This is similar to the commonly-used freely jointed bead-spring FENE
model model used in time-driven MD. The length of the tethers has
been chosen to be $1.1D_{b}$, where $D_{b}$ is the diameter of the
beads %
\footnote{Note that the hard-sphere model rigorously prevents chain crossing
if the tether length is less than $\sqrt{2}D_{b}$ since two tethers
shorter than this length cannot pass through each other without violating
impenetrability.%
}. The implementation of square-well potentials is based on the use
of near-neighbor lists (NNLs) in EDMD, and allows for the specification
of square-well interactions for arbitrary pairs of near neighbors.
In particular, one can specify a minimal $L_{t}^{min}\geq D_{b}$
and maximal distance (tether length) $L_{t}^{max}>L_{t}^{min}$ for
arbitrary pairs of near neighbors %
\footnote{A value $L_{t}^{min}>D_{b}$ can be used to emulate chain rigidity
(i.e., a finite persistence length) by using second nearest-neighbor
interactions between chain beads.%
}.

\section{\label{sec:Algorithmic-Details}Details of Hybrid Algorithm}

In this section the hybrid EDMD/DSMC algorithm, which we name Stochastic
EDMD (SEDMD), is described in detail. Only a brief review of the basic
features of EDMD is given and the focus is on the DSMC component of
the algorithm and the associated changes to the EDMD algorithm described
in detail in Ref. \cite{Event_Driven_HE}. A more general description
of asynchronous event-driven particle algorithms is given in Ref.
\cite{AED_Serial}. 

Asynchronous event-driven (AED) algorithms process a sequence of \emph{events}
(e.g., collisions) in order of increasing event time $t_{e}$. The
time of occurrence of events is predicted and the event is scheduled
to occur by placing it an \emph{event queue}. The simulation iteratively
processes the event at the head of the event queue, possibly scheduling
new events or invalidating old events. One \emph{impending event}
per particle $i$, $1\leq i\leq N$, is scheduled to occur at time
$t_{e}$ with partner $p$ (e.g., another particle $j$). The particle
position $\V{r}_{i}$ and velocity $\V{v}_{i}$ are \emph{only} updated
when an event involving particle $i$ is processed and the time of
last update $t_{i}$ is recorded (we will refer to this procedure
as a \emph{particle update}). We note that traditional \emph{synchronous
time-driven} (STD) algorithms with a \emph{time step} $\D{t}$ are
a trivial variant of the more general AED class. In particular, in
an STD algorithm events occur at equispaced times and each event is
a \emph{time step} requiring an update of all of the particles. The
AED algorithm processes a mixture of events involving single particles
or pairs of particles with time steps that involve the simultaneous
(synchronous) update of a large collection of particles.

Every particle $i$ belongs to a certain specie $s_{i}$. Particles
with species $s_{i}$ and $s_{j}$ may or may not interact with each
other (i.e., they may not be subject to the hard-particle non-overlap
condition). We focus on a system in which a large fraction of the
particles belong to a special specie $s_{DSMC}$ representing DSMC
particles (e.g., solvent molecules). These DSMC particles do not interact
with each other (i.e., they freely pass through each other), but they
do interact with particles of other species. We focus on the case
when the non-DSMC particles are localized in a fraction of the simulation
volume, while the rest of the volume is filled with DSMC particles.
This will enable us to treat the majority of DSMC particles sufficiently
far away from non-DSMC particles more efficiently than those that
may collide with non-DSMC particles.

Before describing the SEDMD algorithm in detail, we discuss the important
issue of efficiently searching for nearby pairs of particles.

\subsection{Near Neighbor Searches}

When predicting the impending event of a given particle $i$, the
time of potential collision between the particle and each of its \emph{neighbors}
(nearby particles) is predicted \cite{Event_Driven_HE,AED_Serial}.
The DSMC algorithm also requires defining neighbor particles, that
is, particles that may collide stochastically during the DSMC collision
step. For efficiency, geometric techniques are needed to make the
number of neighbors of a given particle $O(1)$ instead of $O(N)$.

In SEDMD we use the so-called \emph{linked list cell} (LLC) method
for neighbor searching in both the MD and DSMC components. The simulation
domain is partitioned into $N_{cells}$ cells as close to cubical
as possible. Each particle $i$ stores the cell $c_{i}$ to which
its centroid belongs, and each cell $c$ stores a list $\mathcal{L}_{c}$
of all the particles it contains, as well as the total number of particles
$N_{c}$ in the cell. For a given interaction range, neighbors are
found by traversing the lists of as many neighboring cells as necessary
to ensure that all particles within that interaction range are covered.
In traditional DSMC, only particles within the same cell are considered
neighbors and thus candidates for collision. There are also variants
of DSMC in which particles in nearby cells are included in order to
achieve a non-ideal equation of state \cite{DSMC_Enskog,DSMC_Enskog_Frezzotti}.
A more general implementation would use different cell meshes for
MD and DSMC neighbor searches, however, that would significantly complicate
the implementation.

\subsubsection{Cell Bitmasks}

In addition to the list of particles $\mathcal{L}_{c}$, each cell
$c$ stores a \emph{bitmask} $\mathcal{M}_{c}$ consisting of $N_{\mbox{bits}}>N_{s}+4$
bits (bitfields). These bits may be one (set) or zero (not set) to
indicate certain properties of the cell, specifically, what species
of particles the cell contains, whether the cell is event or time
driven, and to specify boundary conditions. In order to distinguish
the cells that contain non-DSMC particles (i.e., particles of specie
other than $s_{DSMC}$) from those that contain only DSMC particles,
bit $\gamma$ is set if the cell may contain a particle of specie
$\gamma$. The bit is set whenever a particle of specie $\gamma$
is added to the cell, and all of the masks are reset and then re-built
(i.e., refreshed) periodically. When performing a neighbor search
for a particle $i$, cells not containing particles of species that
interact with specie $s_{i}$ are easily found (by OR'ing the cell
masks with a specie mask) and are simply skipped. This speeds up the
processing of DSMC particles since cells containing only DSMC particles
will be skipped without traversing their lists of particles.

For the purposes of the combined MD/DSMC algorithm we will also need
to distinguish those cells that are nearby non-DSMC particles, that
is, that contain particles within the interaction range of some non-DSMC
particle. Such cells will be treated using a fully event-driven (ED)
scheme, while the remaining cells will be treated using a time-driven
or mixed approach. We use one of the bits in the bitmasks, bit $\gamma_{ED}$,
to mark \emph{event-driven (ED) cells} whenever a neighbor search
is performed for a non-DSMC particle. Specifically, bit $\gamma_{ED}$
is set for a given cell whenever the cell is traversed during a neighbor
search for a non-DSMC particle. This scheme correctly masks the cells
by only modifying the neighbor search routines without changing the
rest of the algorithm, at the expense of a small overhead. We also
mark the cells near hard-wall boundaries as ED cells. Cell bitmasks
should be refreshed (rebuilt) periodically so as to prevent the fraction
of ED cells from increasing. As will be seen shortly, it is necessary
to introduce at least one {}``sticky'' bit $\gamma_{st}$ that is
not cleared but rather persists (has memory), and is initialized to
zero (not set) at the beginning of the simulation.

\subsubsection{Near Neighbor Lists}

The cell size should be tailored to the DSMC portion of the algorithm
and can become much smaller than the size of non-DSMC particles. The
LLC method becomes inefficient when the interaction (search) range
becomes significantly larger than the cell size because many cells
need to be traversed. In this case the LLC method can be augmented
with the \emph{near-neighbor list} (NNL) method, and in particular,
the bounding sphere complexes (BSCs) method, as described in detail
for nonspherical hard particles in Ref. \cite{Event_Driven_HE}. We
have implemented the necessary changes to the algorithm to allow the
use of NNLs and BSCs (in addition to LLCs), and we used NNLs in our
simulations of polymer chains in solution. The use of BSCs is not
necessary for efficient simulations of polymer solutions if the size
of the polymer bead is comparable to the size of the cells, which
is the case for the simulations we report. We do not describe the
changes to the algorithm in detail; rather, we only briefly mention
the essential modifications.

For the purposes of DSMC it is important to maintain accurate particle
lists $\mathcal{L}_{c}$ for all cells $c$, so that it is known which
particles are in the same cell (and thus candidates for stochastic
collisions) at any point in time. Therefore, transfers of particles
between cells need to be predicted and processed even though this
is not done in the NNL algorithm described in Ref. \cite{Event_Driven_HE}.
Near-neighbor lists are \emph{only} built and maintained for DSMC
particles that are in event-driven cells (essentially exactly as described
in Ref. \cite{Event_Driven_HE}). For a DSMC particle $i$ that is
not in an ED cell $c_{i}$ we consider the smallest sphere enclosing
cell $c_{i}$ to be the (bounding) neighborhood (see Ref. \cite{AED_Serial})
of particle $i$ and \emph{only} update the (position of the) neighborhood
when the particle moves to another cell. This ensures that neighbor
searches using the NNLs are still exact without the overhead of predicting
and processing \emph{NNL update} events for the majority of the DSMC
particles.

\subsection{\label{Section_EDTD}The SEDMD Algorithm}

We have developed an algorithm that combines time-driven DSMC with
event-driven MD by splitting the particles between ED particles and
TD particles. Roughly speaking, only the particles inside event-driven
cells (i.e., cells for which bit $\gamma_{ED}$ is set) are part of
the AED algorithm. The rest of the particles are DSMC particles that
are not even inserted into the event queue. Instead, they are handled
using a time-driven (TD) algorithm very similar to that used in classical
DSMC. 

It is also possible to implement DSMC as a fully asynchronous event-driven
(AED) algorithm and thus avoid the introduction of an external time
scale through the time step $\D{t}$. The algorithm introduces a novel
type of event we term \emph{stochastic} (DSMC) \emph{collisions},
and it is discussed in more detail in Appendix \ref{Appendix_AEDDSMC}.
Asynchronous processing has a few advantages over the traditional
(synchronous) time-driven approach, notably, no errors due to time
discretization \cite{DSMC_TimeStepError} and improved efficiency
at low collision rates. For high densities (i.e., high collision rates)
we have found that these advantages are outweighed by the (implementation
and run-time) cost of the increased algorithmic complexity. Additionally,
time-driven handling has certain important advantages in addition
to its simplicity, notably, the synchrony of the DSMC portion of the
algorithm allows for parallelization and easy incorporation of algorithmic
alternatives (e.g., multi-particle or multi-cell collisions, adaptive
open boundary conditions, etc.).

The main types of events in the SEDMD algorithm are:

\begin{description}
\item [{Update}] Move particle $i$ to the current simulation time $t$
if $t_{i}<t$.
\item [{Transfer}] Move particle $i$ from one cell to another when it
crosses the boundary between two cells (this may also involve a translation
by a multiple of the lattice vectors when using periodic BCs).
\item [{Hard-core~collision}] Collide a particle $i$ with a boundary
such as a hard wall or another particle $j$ with which it interacts.
\item [{Tether~collision}] Bouncing of a pair of tethered particles in
a polymer chain when the tether stretches (processed exactly like
usual hard-particle collisions \cite{EventDriven_Alder,EDMD_Polymers_Hall}).
\item [{Time~step}] Move all of the time-driven particles by $\D{t}$
and process stochastic collisions between them.
\end{description}
The position $\V{r}_{i}$ and time $t_{i}$ as well as the impending
event prediction of particle $i$ are updated whenever an event involving
the particle is processed.

Both the event-driven and the time-driven DSMC algorithms process
stochastic binary \emph{trial collisions}. Processing a trial collision
consists of randomly and uniformly selecting a pair of DSMC particles
$i$ and $j$ that are in the same cell. For hard spheres in the low-density
limit, the probability of collision for a particular pair $ij$ is
proportional to the relative velocity $v_{ij}^{\mbox{rel}}$, and
therefore the pair $ij$ is accepted with probability $v_{ij}^{\mbox{rel}}/v_{\mbox{rel}}^{\mbox{\mbox{max}}}$.
If a pair is accepted for collision than the velocities of $i$ and
$j$ are updated in a random fashion while preserving energy and momentum
\cite{DSMCReview_Garcia}. If a real collision involving an ED particle
$i$ occurs then that particle is updated to time $t_{TS}$, its previous
event prediction is invalidated (this may involve updating a third-party
particle $k$), and an immediate update event is scheduled for $i$
(and possibly $k$).

It is important to note that the division of the DSMC particles between
ED and TD handling is dynamic and does not necessarily correspond
to the partitioning of the cells into ED and TD cells (based on the
cell bitfield $\gamma_{ED}$). As non-DSMC particles move, time-driven
cells may be masked as event-driven. This does not immediately make
the DSMC particles in such cells event-driven. Rather, time-driven
DSMC particles are moved into the event queue only when a collision
with a non-DSMC particle is scheduled for them, when they move into
a TD cell following a time step, or when restarting the event handling.
Event-driven particles are removed from the event queue when they
undergo cell transfer events into time-driven cells.

\subsubsection{Time Step Events}

The hybrid ED/TD algorithm introduces a new kind of event (not associated
with any particular particle) called a \emph{time step event}. This
event is scheduled to occur at times $t_{TS}=n\D{t}$, where $n\in\mathcal{Z}$
is an integer. When such an event is processed, all of the DSMC particles
not in the event queue are moved %
\footnote{Note that this update may involve moving some particles by less than
$\D{t}$ since the time of the last update for such particles does
not have to be a time step event but could be, for example, a cell
transfer.%
} to time $t_{TS}$ and are then re-sorted into cells (recall that
the ED particles are already correctly sorted into cells). Particles
that change from ED to TD cells and vice-versa are removed or inserted
into the event queue accordingly. Then, in each cell $\Gamma_{c}\D{t}$
trial DSMC collisions are performed, where\begin{equation}
\Gamma_{c}=\frac{N_{C}(N_{C}-1)\sigma v_{max}}{V_{c}}\label{eq:Gamma_c}\end{equation}
is the DSMC collision rate. Here $\sigma=4\pi R_{DSMC}^{2}$ in three
dimensions and $\sigma=4R_{DSMC}$ in two dimensions is the collisional
cross-section, $V_{c}$ is the volume of the cell, and $v_{max}$
is an upper bound for the maximal particle velocity %
\footnote{More precisely, $2v_{max}$ is an upper bound on the maximal relative
velocity between a pair of particles. In our implementation we maintain
the maximal encountered particle velocity $v_{max}$ and update it
after every collision and also reset it periodically.%
}.

In order to ensure correctness of the AED algorithm, a TD particle
must not move by more than a certain distance $\D{l_{max}}$ when
it undertakes a time step. Otherwise, it may overlap with a non-DSMC
particle that could not have anticipated this and scheduled a collision
accordingly. Specifically, recall that the event-driven cells are
marked whenever a neighbor search is performed for a non-DSMC particle.
Our simulation uses \[
\D{l_{max}}=(w_{ED}L_{c}-D_{DSMC})/2,\]
where the masking width $w_{ED}$ is the minimal number of cells covered
by any neighbor search in any direction, $L_{c}$ is the (minimal)
cell length, and $D_{DSMC}$ is the diameter of the DSMC particles.
Any DSMC particle whose velocity exceeds $v_{max}=\D{l_{max}}/\D{t}$
is inserted into the event queue at the end of a time step, and similarly,
any particles that would have been removed from the event queue are
left in the queue if their velocity exceeds the maximum safe velocity.
Typically, only a small (albeit non-zero) fraction of the DSMC particles
falls into this category and the majority of the particles that are
not in ED cells are not in the event queue. In fact, we choose the
time step to be as large as possible while still keeping the number
of dangerously fast DSMC particles negligible. This typically also
ensures that DSMC particles do not jump over cells from one time step
to the next (given that typically $w_{B}=1-2$).

\subsection{\label{Section_OpenBCs}Adaptive Open Boundary Conditions}

In three dimensions, a very large number of solvent particles is required
to fill the simulation domain. The majority of these particles are
far from the polymer chain and they are unlikely to significantly
impact or be impacted by the motion of the polymer chain. It therefore
seems reasonable to approximate the behavior of the solvent particles
sufficiently far away from the region of interest with that of a quasi-equilibrium
ensemble in which the positions of the particles are as in equilibrium
and the velocities follow a local Maxwellian distribution (the mean
of which is equal to the macroscopic local velocity). These particles
do not need to be simulated explicitly, especially for a DSMC liquid
which has no spatial structure (ideal gas). Rather, we can think of
the polymer chain and the surrounding DSMC fluid as being embedded
into an infinite reservoir of DSMC particles which enter and leave
the simulation domain following the appropriate distributions.

Such \emph{open (Grand Canonical) boundary conditions} (BC) are often
used in multi-scale (coupled) simulations. It is not trivial to implement
them when coupling the {}``reservoir'' to an MD simulation, especially
at higher densities. An example of an algorithm that achieves such
a coupling for soft-particle systems is USHER \cite{TetheredPolymer_HybridMD}.
It is also non-trivial to account for the velocity distribution of
the particles entering the simulation domain \cite{DSMC_InflowDistribution},
as would be needed in a purely event-driven algorithm in which particles
are inserted at the surface boundary of the domain. However, the combination
of a partially time-driven algorithm and an unstructured (ideal gas)
DSMC fluid makes it very easy to implement open BCs by inserting DSMC
particles in the cells surrounding the simulation domain only at time-step
events, based on very simple distributions.

\subsubsection{Cell Partitioning}

For the purposes of implementing such non-trivial BCs, we classify
the cells as being \emph{interior, boundary, and external cells}.
Interior cells are those that are in the vicinity of non-DSMC particles,
specifically, cells that are within a window of half-width $w_{int}>w_{ED}$
cells around the centroid of a non-DSMC particle. The interior cells
are divided into event-driven and time-driven and are handled as described
previously. If a boundary or external cell is marked as an event-driven
cell the simulation is aborted with an error, ensuring that ED cells
are always interior. Boundary cells surround the interior cells with
a layer of cells of thickness $w_{B}\geq1$ cells, and they represent
cells in which particles may be inserted during time step events %
\footnote{Since ED cells are never boundary cells such insertions cannot lead
to overlaps with non-DSMC particles.%
}. External cells are non-interior cells that are not explicitly simulated,
rather, they provide a boundary condition around the interior and
boundary cells. This layer must be at least $w_{B}$ cells thick,
and the cells within a layer of $w_{B}$ cells around the simulation
domain (interior together with boundary cells) are marked as both
external and boundary cells. All of the remaining cells are purely
external cells and simply ignored by the simulation. Our implementation
uses bits in the cell bitmasks to mark a cell as being event-driven
(bit $\gamma_{ED}$), boundary (bit $\gamma_{B}$), or external (bit
$\gamma_{P}$). Note that a cell may be a combination of these, for
example, cells near hard walls might be both interior and boundary,
and some cells may be both external and boundary.

\begin{figure}
\includegraphics[width=0.4\paperwidth,keepaspectratio]{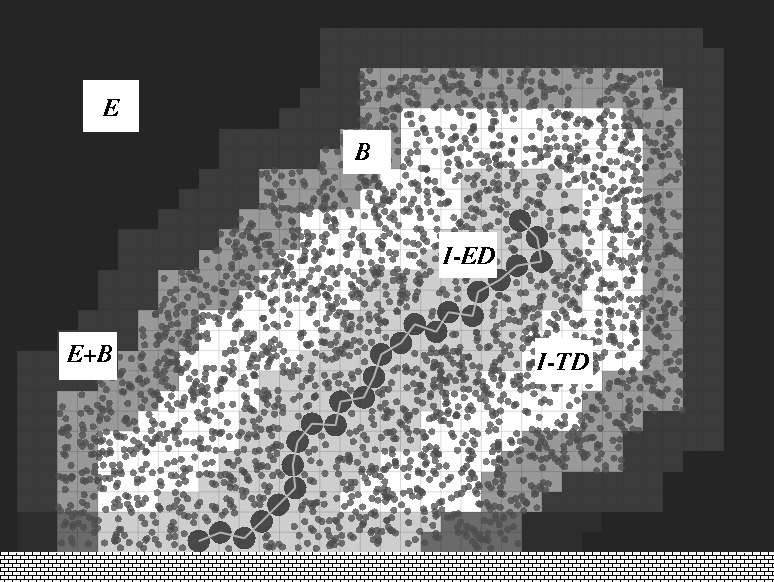}\hspace{0.5cm}\includegraphics[width=0.3\paperwidth,keepaspectratio]{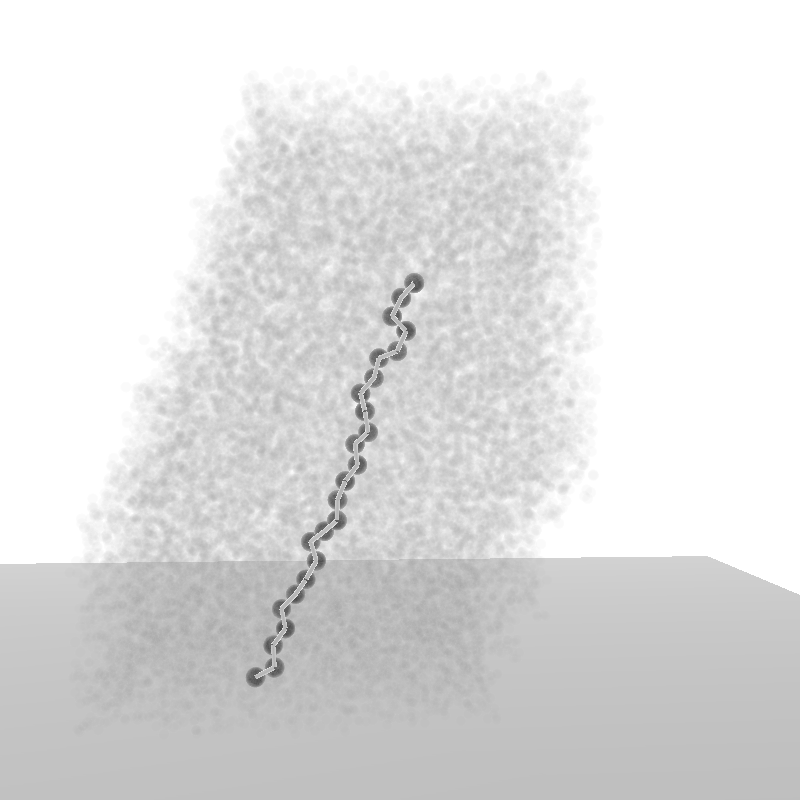}

\caption{\label{TetheredPolymer.partitioning}The partitioning of the domain
into interior (I) {[}either event-driven (ED) or time-driven (TD)],
boundary (B), and external (E) cells in two (left) and three (right)
dimensions for a polymer chain of $25$ beads tethered to a hard wall.
The cells are shaded in different shades of gray and labeled in the
two-dimensional illustration ($w_{ED}=2$, $w_{int}=5$, $w_{B}=2$).
The DSMC particles are also shown.}
\end{figure}

Figure \ref{TetheredPolymer.partitioning} provides an illustration
of this division of the cells for the simulation of a tethered polymer
in two and three dimensions. Note that we do not require that the
domains of interior or non-external cells form a rectangular domain:
The final shapes and even contiguity of such domains depends on the
positions of the non-DSMC particles %
\footnote{If this is not appropriate one can always make the simulation regions
(unions of disjoint) rectangular domains simply by padding with interior
cells.%
}. Our implementation traverses each of the non-DSMC particles in turn
and masks the cells in a window of half-width $w$ cells around the
cell containing the non-DSMC particle as:

\begin{description}
\item [{Interior}] $0\leq w\leq w_{int}$ representing cells where the
non-trivial flow occurs ($w_{int}>w_{ED}$)
\item [{Boundary}] $w_{int}<w\leq w_{int}+2w_{B}$ representing cells where
particles may be inserted or propagated during time step events.
\item [{External}] $w>w_{int}+w_{B}$ representing cells that are not explicitly
simulated but rather only provide appropriate BCs.
\end{description}
The division of the cells into event-driven, interior, boundary and
external cells is rebuilt periodically during the simulation. This
rebuilding may only happen at the beginning of time steps, and requires
a synchronization of all of the particles to the current simulation
time, a complete rebuilding of the cell bitmasks, and finally, a re-initialization
of the event processing. Importantly, particles that are in purely
external cells are removed from the simulation and those that are
in event-driven cells are re-inserted into the event queue scheduled
for an immediate update event. During the process of rebuilding the
cell bitmasks cells that are masked as purely external cells are also
marked with the sticky bit $\gamma_{s}$. This indicates that these
cells need to be re-filled with particles later if they enter the
simulation domain again (due to the motion of the non-DSMC particles).
Once the cell bitmasks are rebuilt, a time step event is executed
as described next.

\subsubsection{Reservoir Particles}

At the beginning of a time step event, after possibly rebuilding the
cell masks, the time-driven DSMC particles are propagated as usual.
If there are external cells, (trial) \emph{reservoir particles} are
then inserted into the cells that are both external and boundary,
and also in cells whose sticky bit $\gamma_{s}$ is set (i.e., cells
that have not yet been filled with particles), after which the bit
$\gamma_{s}$ is reset. The use of the sticky bit to mark such cells
ensures that subsequent rebuilding of the masks will not erase the
flag (the sticky bit is \emph{only} reset once the cell is filled
with particles). The trial particles are thought to be at a time $t-\D{t}$,
and are propagated by a time step $\D{t}$ to the current simulation
time. Only those particles that move into a non-external cell are
accepted and converted into real particles. If the acceptance would
insert the particle into a non-boundary cell (i.e., the particle moved
by at least $w_{B}$ cells), the insertion is rejected and a count
of the number of rejected particles reported to aid in choosing $w_{B}$
sufficiently large so as to ensure that the tails of the velocity
distribution are not truncated (in our experience $w_{B}=2$ suffices
for reasonable choices of $\D{t}$). Following the insertion of reservoir
particles stochastic collisions are processed in each cell as usual. 

For improved efficiency, it is possible to replace the volume-based
particle reservoir with a surface reservoir, and insert particles
only at the surface of the simulation domain \cite{DSMC_InflowDistribution}.
However, we have not implemented such an approach since the boundary
handling is not critical for the overall efficiency.

\subsubsection{Boundary Conditions}

In our current implementation the reservoir particles follow simple
local-equilibrium ideal gas distributions. The number of particles
to insert in a given cell $c$ is chosen from a Poisson distribution
with the appropriate density, the positions are uniformly distributed
inside the cell, and the velocities are drawn from a biased (local)
Maxwellian distribution. The mean velocity $\V{v}_{M}$ and temperature
$T_{M}$ for the local Maxwellian are chosen according to the specified
boundary conditions (presently only uniform linear gradients are implemented).
For example, if a uniform shear in the $xy$ plane is to be applied,
$\V{v}_{M}=\gamma y_{c}\hat{x}$, where $y_{c}$ is the $y$ position
of the centroid of the cell and $\gamma$ is the shear rate. Using
such biased local insertions allows one to specify a variety of boundary
conditions (for example, a free polymer chain in \emph{unbounded}
shear flow) without resorting to hard-wall boundaries or complicating
Lee-Edwards conditions.

It should be noted that in principle we should not use a local Maxwellian
velocity distribution for a system that is not in equilibrium. In
particular, for small velocity, temperature, and density gradients
the Chapman-Enskog distribution is the appropriate one to use in order
to avoid artifacts near the open boundaries at length scales comparable
to the mean free path $\lambda$ \cite{AMAR_DSMC}. We judge these
effects to be insignificant in our simulations since our boundary
conditions are fixed externally and are thus not affected by the possible
small artifacts induced in the DSMC fluid flow, and since $\lambda$
is small.

In the future, we plan to replace the particle reservoir with a PDE-based
(Navier-Stokes) simulation coupled to the DSMC/MD one. Such a flux-preserving
coupling has been implemented in the past for coupled DSMC/Euler hydrodynamic
simulations \cite{AMAR_DSMC,AMAR_DSMC_SAMRAI}. It is however important
for the coupling to also correctly couple fluctuations. This requires
the use of \emph{fluctuating hydrodynamics} in the coupled domain.
Such solvers and associated coupling techniques are only now being
developed \cite{FluctuatingHydro_Garcia,FluctuatingHydro_AMAR}.

\subsection{Further Technical Details}

In this section we discuss several technical details of the SEDMD
algorithm such as hard-wall boundary conditions and the choice of
DSMC parameters.

\subsubsection{\label{Section_NoSlip}Slip and Stick Boundary Conditions}

We have already discussed the open boundary conditions and their use
to specify a variety of {}``far-field'' flow patterns. Additionally,
there can also be hard-wall boundaries, i.e., flat impenetrable surfaces.
These surfaces can have a velocity of their own and here we discuss
how particles reflect from such walls in the frame that moves with
the hard wall. Regardless of the details of particle reflections,
the total change in linear momentum of all the particles colliding
with a hard wall can be used to estimate the friction (drag) force
acting on the wall due to the flow. This can give reliable and quick
estimates of the viscosity of a DSMC fluid, for example. We use the
classical no-slip BCs (i.e., zero normal and parallel velocity) for
smooth hard-wall surfaces. Molecular simulations have found some slip;
however, at length-scales significantly larger than the mean free
path and/or the typical surface roughness one may assume no-slip boundaries
if the hard-wall boundary position is corrected by a slip length $L_{slip}$
\cite{Microfluidics_Review}.

Our simulations of tethered polymers use \emph{thermal walls} (kept
at $kT=1$) \cite{DSMCReview_Garcia} to implement no-slip hard walls
at the boundaries of the simulation cell. Following the collision
of a particle with such a wall, the particle velocity is completely
randomized and drawn from a half Maxwell-Boltzmann distribution (other
biased distributions may be used as appropriate). This automatically
ensures a zero mean velocity at the wall boundary and also acts as
a thermostat keeping the temperature constant even in the presence
of shear heating. No-slip boundaries can also be implemented using
(athermal) \emph{rough walls} which reflect incoming particles with
velocity that is the exact opposite of the incoming velocity \cite{DSMC_MPCD_CylinderFlow}.
Similarly, slip boundary conditions (zero normal velocity) can be
trivially implemented by using \emph{specular walls} that only reverse
the normal component of the velocity (relative to the wall). A mixture
of the two can be used to implement partially rough walls, for example,
a roughness parameter $0\leq r_{w}\leq1$ can be used as the probability
of randomly selecting a rough versus a specular collision.

Similar considerations apply to the boundary conditions at the interface
of a hard particle such as a polymer bead. Most particle-based methods
developed for the simulation of particle suspensions consider the
solvent particles as point particles for simplicity, and only MD or
certain boundary discretization schemes \cite{SuspensionsDSMC_Reflection}
resolve the actual solvent-solute interface. Specular BCs are typical
of MD simulations and assume perfectly conservative collisions (i.e.,
both linear momentum and energy are conserved). However, if the polymer
beads are themselves composed of many atoms, they will act as a partially
thermal (and rough) wall and energy will not be conserved exactly. 

In the simulations reported here we have used rough walls for collisions
between DSMC and non-DSMC particles. This emulates a non-stick boundary
condition at the surface of the polymer beads. Using specular (slip)
conditions lowers the friction coefficient %
\footnote{The Stokes friction force has a coefficient of $4\pi$ for slip BCs
instead of the well-known $6\pi$ for no-slip BCs.%
}, but does not appear to qualitatively affect the behavior of tethered
polymers.

\subsubsection{\label{Section_ConstantPressure}Constant Pressure Flows}

We note briefly on our implementation of constant pressure boundary
conditions, as used to simulate flow through open pipes. A constant
pressure is typically emulated in particle simulations via a constant
acceleration $\V{a}$ for the DSMC particles \cite{DSMC_PlanePoiseuille}
together with periodic BCs along the flow (acceleration) direction.
In time-driven algorithms, one simply increments the velocity of every
particle by $\V{a}\D{t}$ and the position by $\V{a}\D{t}^{2}/2$
at each time-step (before processing DSMC collisions). In SEDMD this
is not easily implemented, since the trajectory of the DSMC particles
becomes parabolic instead of linear and exact collision prediction
between the DSMC and the non-DSMC particles is complicated. We have
opted to implement constant pressure BCs by using a periodic delta-function
forcing on the DSMC particles. Specifically, the velocities of all
DSMC particles are incremented %
\footnote{Recall that the event prediction for any ED particle $i$ whose velocity
is changed must be updated, typically by scheduling an immediate update
event. %
} at the beginning of each time step by $\V{a}\D{t}$, and then stochastic
collisions are processed.

\subsubsection{\label{Section_Gamma_c}Choice of DSMC Collision Frequency}

The viscosity of the DSMC fluid is determined by the choice of collision
frequency $\Gamma_{c}$ and cell size $L_{c}$. Classical DSMC wisdom
\cite{DSMCReview_Garcia} is that cell size should be smaller than
the mean free path, $L_{c}\ll\lambda$, but large enough to contain
on the order of $N_{c}\approx20$ particles (in three dimensions).
It is obvious that both of these conditions cannot be satisfied for
denser liquids, where $\lambda$ is only a fraction of the particle
size. It is now well-known that it is not necessary to have many particles
per cell, so long as in Eq. (\ref{eq:Gamma_c}) we use $N_{c}(N_{c}-1)$
instead of the traditional (but wrong) $N_{c}^{2}$. Coupled with
the Poisson distribution of $N_{c}$ this gives a constant average
total collision rate. However, using very small cells leads to very
large variability of collision rates from cell to cell and thus spatial
localization of momentum transfer during each time step. Namely, with
very small cells one rarely has two particles and thus most of the
collisions will occur in the few cells that happen to be densely populated.

We have aimed at trying to mimic what would happen in an MD simulation
in the DSMC one. In an MD simulation particles collide if their distance
is equal to the particle diameter $D$. Therefore, we have aimed at
keeping the cell size at a couple of diameters, $L_{c}\approx2D$.
At typical hard-sphere liquid densities this leads to $N_{c}\approx5-10$,
which seems appropriate in that it allows enough collision partners
for most of the particles but still localized the momentum transfer
sufficiently. For very small mean free paths DSMC does not distinguish
velocity gradients at length scales smaller than the cell size and,
in a long-time average sense, localizes the velocity gradients at
cell interfaces \cite{DSMC_CellSizeError}. We will assume that, for
problems of interest to us, the structure of the fluid and flow at
length-scales comparable to $D$ (and thus $L_{c}$) is unimportant,
and verify this by explicit comparisons to MD.

When the cell size is chosen such that $N_{c}\approx5-10$ and the
time step is reasonable, $\D{t}\approx(0.1-0.2)L_{c}/\bar{v}$, Eq.
(\ref{eq:Gamma_c}) gives collision frequencies that are sufficiently
high so that almost all particles suffer at least one collision every
time step, and typically more than one collision. The effect of such
repeated collisions is to completely thermalize the flow to a local
equilibrium (i.e., local Maxwellian), and we have observed that further
increasing the collision frequency does not change the effective viscosity
(we do not have a theoretical understanding of this behavior \cite{DSMC_CellSizeError}).
For greater efficiency, we have chosen to use the lowest collision
rate (for a given timestep) that still achieves a viscosity that is
as high as using a very high collision rate. We find that this is
typically achieved when each particle suffers about half a collision
or one collision each timestep \cite{DSMC_GammaBounds}. Appendix
\ref{Appendix_TRMC} describes some multi-particle collision variants
that may be more appropriate under different conditions.

\subsubsection{\label{Section_noHI}DSMC without Hydrodynamics}

The solvent exerts three primary effects on polymers in flow: (1)
stochastic forces due to fluctuations in the fluid (leading to Brownian-like
motion), (2) (local) frictional resistance to bead motion (usually
assumed to follow Stokes law), and (3) hydrodynamic interactions between
the beads due to perturbations of the flow field by the motion of
the beads. Brownian dynamics, the most common method for simulating
the behavior of polymers in flow, coordinates the first two effects
via the fluctuation-dissipation theorem, and potentially adds the
third one via approximations based on the Oseen tensor (neglecting
the possibility of large changes to the flow field due to the moving
beads). By turning off local momentum conservation one can eliminate
all hydrodynamic interactions, and thus test the importance of the
coupling between polymer motion and flow.

Yeoman's \emph{et al.} \cite{PolymerCollapse_Yeomans,ShearThinning_Yeomans}
have implemented a no-hydrodynamics variant of the MPCD algorithm
by randomly exchanging the velocities between all particles at each
time step (thus preserving momentum and energy globally, but not locally).
In the presence of a background flow, such as shear, only the components
of the velocities relative to the background flow are exchanged. We
have implemented a no-hydrodynamics variant of DSMC by neglecting
momentum conservation in the usual stochastic binary collisions %
\footnote{In this implementation switching hydrodynamics off becomes an alternative
branch localized in the binary collision routine and the algorithm
is otherwise unchanged.%
}. Specifically, if particles $A$ and $B$ collide, the post-collisional
velocity of $A$ is set to be the same magnitude as that of $B$ but
with a random orientation, and vice versa (this conserves energy but
\emph{not} momentum). If the boundary conditions specify a background
flow such as a uniform shear the flow velocity is evaluated at the
center of the DSMC cell and the collisions are performed in the frame
moving with that velocity. This forces the average velocity profile
to be as specified by the boundary conditions, but does not allow
for perturbations to that profile due to hydrodynamic effects.

\section{\label{Section_Performance}Performance Improvement}

It is, of course, expected that the DSMC algorithm will give a performance
improvement over MD. However, to make an impact on real-world problems
this performance gain must be an order of magnitude or more improvement.
Indeed, we find that SEDMD with adaptive boundary conditions can be
up to two hundred times faster than EDMD under certain conditions.
Note also that it is well-known that EDMD is already significantly
faster than TDMD (depending on the density), although such a comparison
is somewhat unfair since the hard-core interaction potentials are
very simple by design. It is important to note that this algorithm
is serial, and we do not consider or use any parallelization. Because
of the inherent simplicity and thus efficiency of the algorithm, however,
it is possible to study time scales and system sizes as large or larger
than parallel simulations described in the literature so far. The
combined time-driven DSMC with event-driven MD algorithm can be parallelized
using traditional techniques from TDMD if proper domain partitioning
is constructed, so that each event-driven region is processed by a
single processor %
\footnote{Achieving good load balancing will be easiest for systems containing
multiple polymer chains.%
}.

As model problem we study a tethered polymer in three dimensions.
The solvent density was chosen to be typical of a moderately dense
hard-sphere liquid. The performance and optimal choice of parameters
depends heavily on the size of the beads relative to the size of the
solvent particles for both MD and the hybrid algorithm. Realistically,
beads (meant to represent a Kuhn segment) should be larger than the
solvent molecules%
\footnote{For example, in Ref. \cite{FluctuatingHydroMD_Coveney} an appropriate
bead size for polyethylene is estimated at $1.5nm$, and for DNA (a
much stiffer molecule with large persistence length) at $40nm$.%
}. This of course dramatically increases the computational requirements
due to the increase in the number of solvent particles (and also makes
neighbor searching more costly). For this reason, most MD simulations
reported in the literature use solvent particles that are equivalent,
except for the chain connectivity, to the solute particles.

Our first test problem is for a chain of $25$ large beads, each about
$10$ times larger (in volume and in mass) than the solvent particles,
in a box of size $2\times1.25\times1.25$ polymer lengths, for a total
of about $N=2.3\times10^{5}$ particles. For the DSMC simulations,
we did not use BSCs (bounding sphere complexes \cite{Event_Driven_HE}),
and therefore the neighbor search had to include next-nearest neighbor
cells as well (i.e., $w_{ED}=2$). For the corresponding MD simulations,
BSCs were used. Under these conditions, DSMC outperformed MD by a
factor of $35$. If adaptive open BCs were used with $w_{int}=5$,
giving about $N=3.2\times10^{4}$ particles (the exact number changes
with polymer conformation), the speedup was $180$. While this may
seem an unfair comparison, it is important to point out that we do
not even know how to implement an adaptive simulation domain in pure
EDMD. 

The second test problem was for a chain of $30$ beads which were
identical to the solvent particles, except for the added chain tethers.
The number of particles in the simulation cell was thus much smaller,
$N=4.8\times10^{4}$, and $w_{ED}=1$. Adaptive BCs with $w_{int}=5$
reduce the simulation domain to $N=2.2\times10^{4}$ particles. For
these parameters DSMC with adaptive BCs was about $30$ times faster
than full MD. Table \ref{PerformanceGains} summarizes the large performance
gains of SEDMD relative to traditional EDMD.

\begin{table}
\begin{tabular}{|c|c|c|}
\hline 
&
Standard BCs&
Adaptive BCs\tabularnewline
\hline
\hline 
Large beads&
35&
180\tabularnewline
\hline 
Small beads&
20&
30\tabularnewline
\hline
\end{tabular}

\caption{\label{PerformanceGains}Performance gains of SEDMD relative to EDMD
for a typical tethered polymer simulation.}
\end{table}

One of the fundamental problems with multi-scale modeling is that
typically the majority of the simulation time is spent in the finest
model since it is difficult to match the time scales of the coupled
components \cite{Trebotich_Penalty}. For example, MD simulations
are so expensive that coupling them to almost any meso- or macro-scopic
solver leads to simulation times limited by that of MD simulations
(albeit of a much smaller system). By virtue of the fast microscopic
algorithm (EDMD instead of TDMD) and the efficient coupling, our method
spends comparable amounts of computation on the solute (and immediately
surrounding solvent) and the solvent particles. For the DSMC run with
adaptive open BCs and large beads, about $50\%$ of the time was spent
in manipulation of near neighbor lists. Most of the remaining time
was spent inside the routine that takes a DSMC timestep, and actual
processing of DSMC collisions (both trial and real) occupied about
$20\%$ of the computation time. For small beads, the majority of
the time, $80\%$, was spent in the DSMC time-step routine, and processing
of DSMC binary collisions occupied about $35\%$ of the computation
time.

\section{\label{sec:Tethered-Polymer}Tethered Polymer in Shear Flow}

In this section results are presented for a tethered polymer chain
in uniform shear in three dimensions. The linear chain is in a good
solvent and is attached at one end to a hard wall, as represented
by the plane $y=0$. A linear velocity profile $v=\gamma y\hat{x}$
along the $x$ axis is imposed sufficiently far from the chain. This
problem was first studied experimentally by Doyle \emph{et al.} \cite{TetheredPolymer_Experiment_PRL}
and since then numerous computational studies have investigated various
aspects of the problem \cite{TetheredPolymer_HybridMD,TetheredPolymer_FullMD,TetheredPolymer_Cyclic_PRL,TetheredPolymer_Cyclic_AIP,TetheredDNA_FullMD}.
We will focus on the dynamics of the chain at low to medium flow rates
(Weissenberg numbers) because we wanted to verify that our polymer
and solvent model can correctly reproduce non-trivial dynamics.

\subsection{Background}

The properties of a linear polymer in shear flow can be related to
the dimensionless Weissenberg number $\Wi=\gamma\tau_{0}$, where
$\tau_{0}=\tau(\gamma=0)$ is the relaxation time of the polymer chain
when there is no shear. When $\Wi<1$ the flow barely affects the
polymer, contrary to when $\Wi>1$. Different models have given similar
properties for the same Weissenberg number.

The original experimental study of tethered polymers \cite{TetheredPolymer_Experiment_PRL}
observed what was termed {}``cyclic dynamics'' of the chains. Specifically,
the following cycle was proposed. When the polymer moves too far from
the wall, presumably by an unusual fluctuation, it experiences a stronger
flow and is stretched. A torque develops that then pushes the chain
closer to the wall, where it can contract again due to the weaker
flow near the wall. The cycle then repeats. Experiments \cite{TetheredPolymer_Experiment_PRL}
did not identify clear periodicity of this motion. Subsequent computational
studies have looked for such a characteristic period for this cycling
motion.

The MD study in Ref. \cite{TetheredPolymer_FullMD} examined the cross-correlation
function $C_{X\phi}(t)$, where $X$ measures the extension of the
polymer along the flow, and $\phi$ measures the angle of the chain
with respect to the hard wall. No exact definitions of $X$ or $\phi$
were given even though there are several possibilities. One can use
the difference between the maximal and the minimal bead positions
as a measure of the extension along a given axes. Optionally, one
can simply use the maximal position, or one can use the position of
the last bead. Similarly, the angle of the polymer can be based on
a linear fit to the shape of the chain, on the position of the center
of mass, the asymmetry of the gyration tensor \cite{PolymerTumbling_PRL},
or the position of the last bead. We have examined various choices
and have found little qualitative difference between the different
choices. We have found the position of the end bead $\V{r}_{N_{b}}=(x,y,z)$
to be the best option and will also measure the angle $\phi=\tan^{-1}(y/x)$.

The authors of Ref. \cite{TetheredPolymer_FullMD} found that $C_{x\phi}(t)$
develops a peak at positive time $t^{*}$ for sufficiently large $\Wi$
numbers ($\Wi>10$). This was interpreted as supporting the existence
of a critical Weissenberg number $\Wi$ where the flow effect on the
polymer dynamics changes qualitatively. It was also found that $t^{*}$
decreases with increasing $\Wi$ and the height of the peak increases.
It is important to note that $t^{*}$ was found to be comparable to
the relaxation time of the polymer $\tau_{0}$. Additionally, the
internal relaxation time $\tau$ was found to decrease with increasing
$\Wi$, in agreement with theoretical predictions.

A subsequent study which used a hybrid MD/CFD model, and also a (free-draining)
Brownian dynamics model, claimed to observe periodic oscillations
in the cross-correlation function between the extensions along the
flow and along the shear direction (i.e., perpendicular to the wall),
$C_{xy}(t)$ \cite{TetheredPolymer_Cyclic_PRL,TetheredPolymer_Cyclic_AIP}.
However, the period of oscillation was found to be an order of magnitude
larger than the internal relaxation time, as revealed by a small peak
in the power spectral density $PSD_{xy}(f)$ of $C_{xy}(t)$. A similar
claim was made in Ref. \cite{PolymerTumbling_PRL} based on $PSD_{\phi\phi}$
of the polymer angle autocorrelation function %
\footnote{The PSD is equivalent to the Fourier spectrum power of the angle trace
$\phi(t)$ based on the convolution theorem.%
} $C_{\phi\phi}(t)$ for both a free polymer in unbounded shear flow
and a tethered polymer in shear flow. No results for the short-time
cross-correlation functions were reported in either of these studies
making it difficult to reconcile the results obtained from PSDs with
those in Ref. \cite{TetheredPolymer_FullMD}.

Most experimental and computational studies of the dynamics of polymers
in shear flow have been for free chains in unbounded flow \cite{PolymerTumbling_Review}.
In that problem, for $\Wi>1$, it is possible to identify a well-defined
{}``tumbling'' event as the polymer rotates. The frequency of such
tumbling times can be measured by visual inspection and have been
compared to the computed location of the peak in the PSDs \cite{PolymerTumbling_PRL,PolymerTumbling_PDFs}.
The good match has thus been taken as an indicator that PSDs peaks
can be used to determine characteristic tumbling times and the same
methodology has been applied to a tethered polymer as well. However,
for the case of a tethered chain it is not easy to identify a periodic
event such as a specific rare fluctuation. Therefore, it is not surprising
that we do not confirm the existence of a characteristic time that
is an order of magnitude larger than the internal relaxation time.
One must here distinguish between {}``cyclic'' (repetitive) events
and periodic events. A Poisson time process of rate $\Gamma$ has
a well-defined time scale $\Gamma^{-1}$, however, the occurrence
of such events is not periodic; the delay between successive events
is exponentially-distributed. In Ref. \cite{PolymerTumbling_PDFs}
such an exponential distribution is proposed even for the delay between
successive tumbling events for a free chain in unbounded flow. The
PSD of such a process is expected to be that of white noise (i.e.,
flat) for frequencies small compared to $\Gamma$, and typically a
power-law decay for larger frequencies (gray noise). The occurrence
and shape of any local maxima (peaks) or frequencies comparable to
$\Gamma$ depends on the exact nature of the correlations at that
time scale.

\subsection{Model Parameters}

As explained in \ref{Section_Performance}, we have made several runs
for different polymer lengths and also bead sizes. One set of runs
used either $N_{b}=25$ or $50$ large beads each about 10 times larger
than a solvent particle, using DSMC with or without hydrodynamics
(see Section \ref{Section_noHI}) for the solvent. Another set of
runs used either $N_{b}=30$ or $60$ small beads each identical to
a solvent particle, using DSMC or pure MD for the solvent. The beads
were rough in the sense that no-slip conditions were applied for the
solvent-solute interface (see Section \ref{Section_NoSlip}). All
of the runs used open boundary conditions (see Section \ref{Section_OpenBCs}),
and the typical half-width of the interior region was $w_{int}=5$
or $w_{int}=7$ cells around the polymer chain. The difference in
the results (such as relaxation times) between these runs and runs
using $w_{int}=10$ or runs using periodic BCs were negligible for
the chain sizes we studied %
\footnote{It is expected that using a small $w_{int}$ would truncate the (long-ranged)
hydrodynamic interactions and thus increase the relaxation time. We
observe such effects for the $N_{b}=50$ chains, however, the effect
is too small compared to the statistical errors to be accurately quantified.%
}. The solvent was a hard-sphere MD or DSMC fluid with volume fraction
$\phi\approx0.25-0.30$, which corresponds to a moderately dense liquid
(the melting point is $\phi_{m}\approx0.49$). The $N_{b}=30$ runs
were run for $T\approx6000\tau_{0}$ with $w_{int}=7$, and such a
run takes about 6 days on a single 2.4GHz Dual-Core AMD Opteron processor.
Even for such long runs the statistical errors due to the strong fluctuations
in the polymer conformations are large, especially for correlation
functions at long time lags $t>\tau$.

\subsection{Relaxation Times}

\begin{figure}
\includegraphics[width=0.9\columnwidth,keepaspectratio]{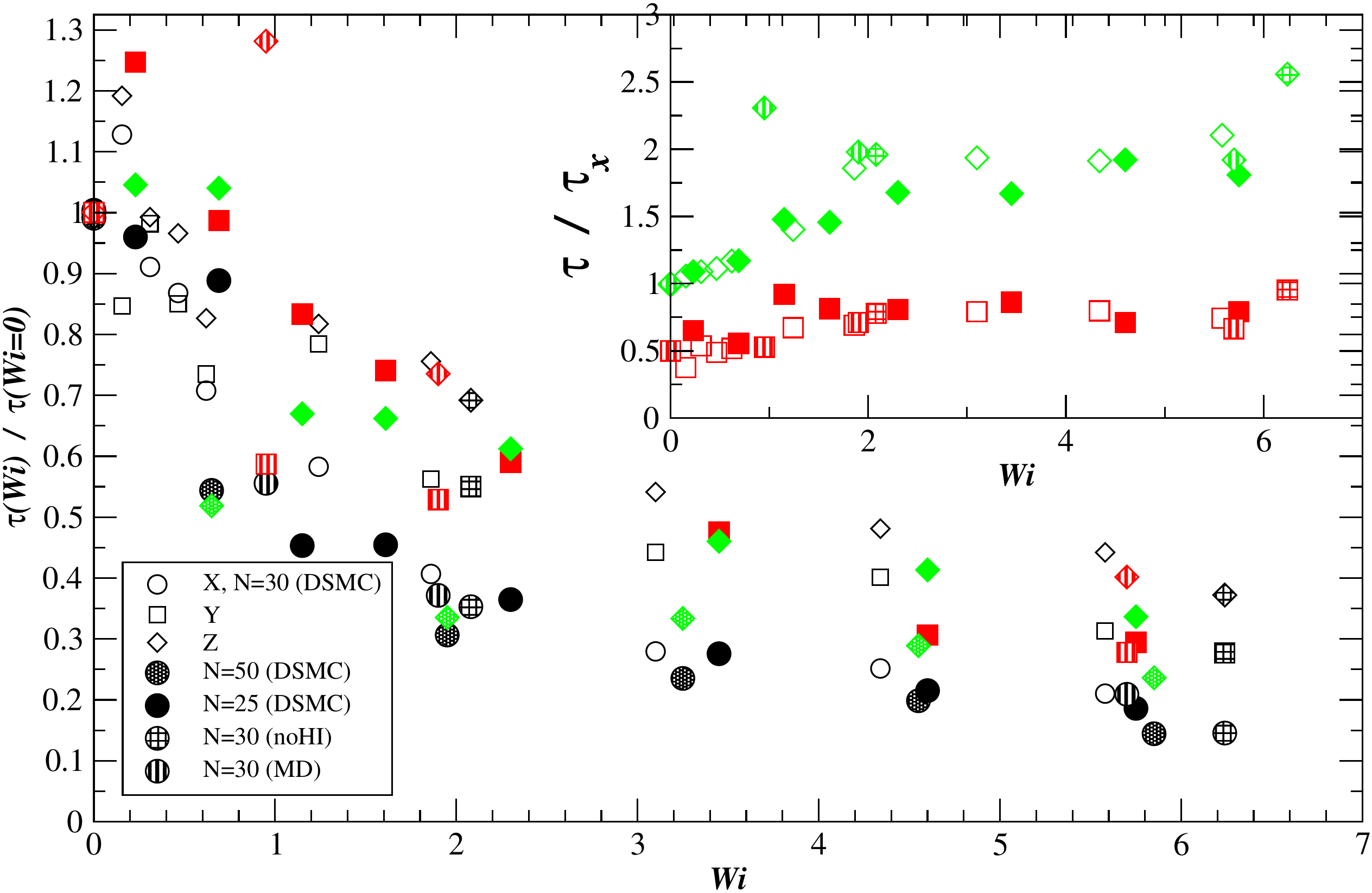}

\caption{\label{tau_Wi.XYZ}Dependence of the relaxation times of the different
components of the end-to-end displacement vector on the Weissenberg
number. The relaxation times have been renormalized to equal unity
for $\Wi=0$ for direct comparison. For each $\Wi$, $\tau_{x}$ is
shown with circles, $\tau_{y}$ with squares, and $\tau_{z}$ with
diamonds. Different textures of the symbols are used for the different
models, as indicated in the legend. The inset shows $\tau_{y}/\tau_{x}$
and $\tau_{z}/\tau_{x}$ for the different runs.}
\end{figure}

The \emph{relaxation time} of the polymer $\tau$ is well-defined
only for linear models. It is often measured by fitting an exponential
to the autocorrelation function of the end-to-end vector $\V{r}_{end}(t)=\V{r}_{N_{b}}-\V{r}_{1}$,
where $\V{r}_{i}$ denotes the position of the $i$-th bead \cite{PolymerDynamics_Review}.
We will separately consider the different components of the end-to-end
vector $\V{r}_{end}=(x,y,z)$ and fit an exponential %
\footnote{The initial relaxation of the various auto-correlation functions $C(t)$
is faster than exponential, and the statistical error at longer times
is large even for long runs. We therefore fit the exponentials to
the portion of the curves at small times, when $0.2\leq C(t)\leq0.8$.
The fits are not perfect and there are large statistical errors depending
on the length of the run and the number of samples used to average
$C(t)$, and the relaxation times (and thus Weissenberg numbers) we
quote should be taken as approximate.%
} to the $C_{xx}$, $C_{yy}$ and $C_{zz}$ auto-correlations functions
to obtain the relaxation times $\tau_{x}$, $\tau_{y}$ and $\tau_{z}$
as a function of $\Wi$. We find that $\tau_{z}$ is always the largest,
especially for large $\Wi$ (for $\Wi=0$, $\tau_{z}=\tau_{x}$ by
symmetry), and $\tau_{y}$ is always smaller by at least a factor
of two %
\footnote{This is because of the constraint that the polymer chain must be above
the plane $y=0$ at all times, which reduces the available configuration
space.%
}, even for $\Wi=0$, as illustrated in the inset in Fig. \ref{tau_Wi.XYZ}.
We take $\tau_{0}=\tau_{x}(\Wi=0)=\tau_{z}(\Wi=0)$ as the definition
of the polymer relaxation time.

The relaxation times we observe for $\Wi=0$ are consistent with what
is predicted from theoretical considerations, $\tau\approx0.9\eta b^{3}N_{b}^{1.8}/kT$,
where $\eta$ is the viscosity and $b$ is the effective bead radius.
Using the viscosity (based on Enskog theory) of the MD liquid %
\footnote{Direct measurements of the viscosity of the DSMC liquid show that
it has viscosity rather close to that of the corresponding MD liquid
for the specific parameters we use.%
} and the tether length as $b$, we calculated $\tau\approx19$ for
the case of $N_{b}=25$ with large beads, to be compared to the numerical
results from DSMC $\tau=25\pm5$. The MD runs for the case of large
beads are not long enough to determine the relaxation time accurately.
We expect that the difference between MD and DSMC will become more
pronounced for smaller beads, and indeed, for $N_{b}=30$ we obtain
$\tau_{MD}\approx3\tau_{DSMC}$. Turning hydrodynamics off in DSMC
extends the relaxation times (and also the collapse times for an initially
stretched polymer) by a factor of $3-5$, as already observed using
MPCD \cite{PolymerCollapse_Yeomans} and as predicted by Zimm theory
\footnote{It is difficult to directly compare DSMC with and without hydrodynamics
since switching hydrodynamics off, in our model, affects the friction
force between the beads and the solvent. This is unlike the models
were the friction force is an added phenomenological term that has
an adjustable coefficient. %
}. Figure \ref{tau_Wi.XYZ} illustrates the dependence of $\tau_{x}(\Wi)/\tau_{x}(\Wi=0)$
on $\Wi$, and similarly for the $y$ and $z$ directions. Quantitatively
similar (but not identical) results are observed independently of
the details of the polymer model and even the existence of hydrodynamic
relaxations.

\subsection{Cyclic Dynamics}

\begin{figure*}
\includegraphics[width=0.95\textwidth,keepaspectratio]{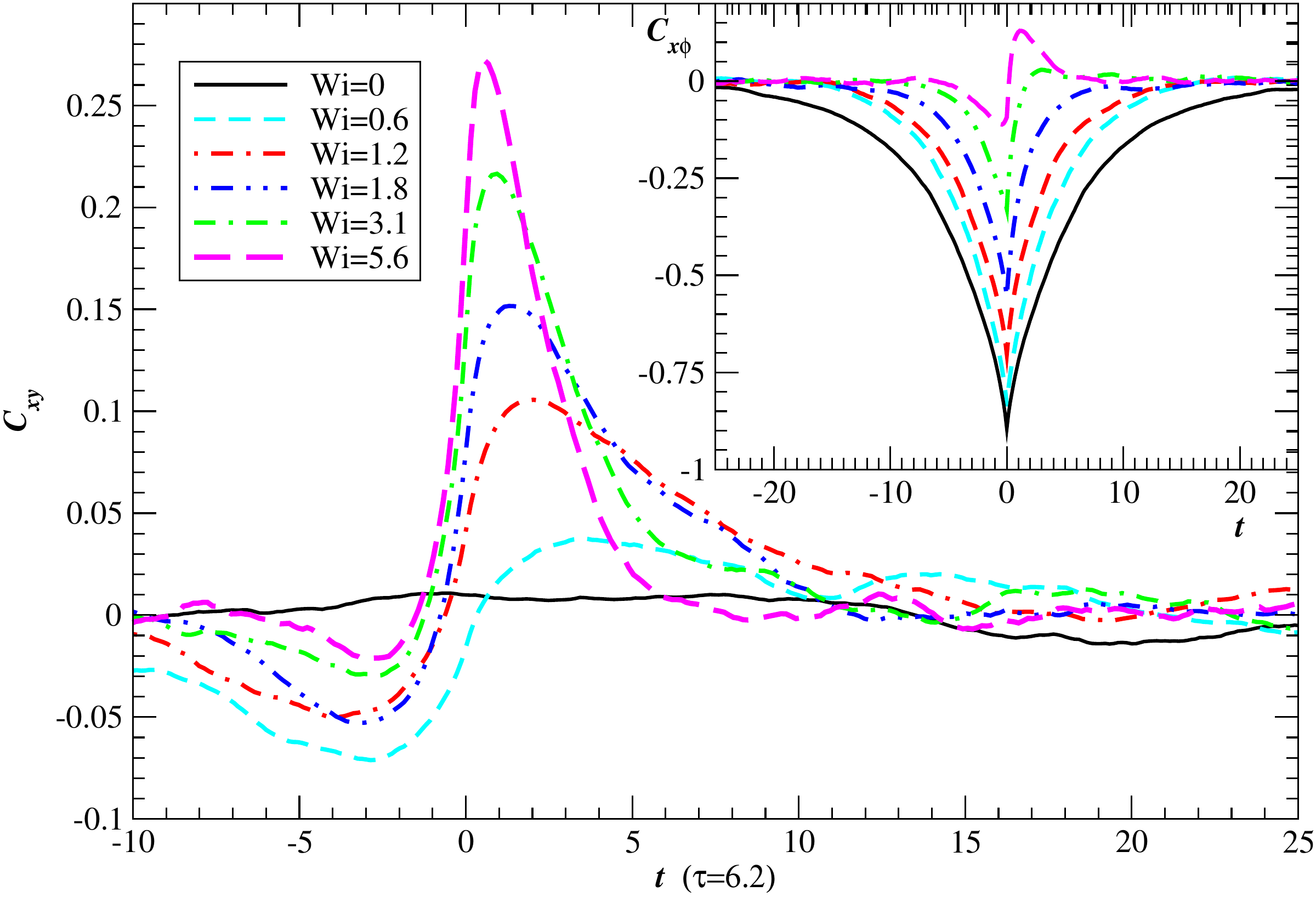}

\caption{\label{C_xy.N=3D30}Cross correlation function $C_{xy}(t)$ for chains
of $N_{b}=30$ small beads in a DSMC solvent at different shear rates.
The inset shows the corresponding $C_{x\phi}(t)$ for comparison with
the soft-particle MD results in Ref. \cite{TetheredPolymer_FullMD}.
Peaks are visible in $C_{xy}(t)$ at all $\Wi>0$, but are obscured
in $C_{x\phi}(t)$ due to the large negative dip at the origin for
$\Wi=0$. There are large statistical errors at small $\Wi$ making
it difficult to identify the peaks.}
\end{figure*}

We now turn our attention to cross-correlations between polymer extensions
in the $x$ and $y$ directions. We have found that the cross-correlations
lags are most visible in the $x$ and $y$ positions of the last bead,
$C_{xy}(t)$. Our results for $C_{xy}(t)$ are shown in Fig. \ref{C_xy.N=3D30},
along with $C_{x\phi}(t)$ as an inset. The results for $C_{x\phi}(t)$
compare well with those in Ref. \cite{TetheredPolymer_FullMD}, although
we see the secondary peak developing at somewhat lower $\Wi$. We
do not see any evidence for the existence of a critical $\Wi$: There
are peaks at both positive and negative time in $C_{xy}(t)$ for all
$\Wi$. Some cross-correlations, such as $C_{x\phi}(t)$, have a large
positive or negative cusp at the origin at $\Wi=0$ and it is this
cusp that masks the peaks at non-zero lags for small $\Wi$. 

\begin{figure}
\includegraphics[width=0.75\columnwidth,keepaspectratio]{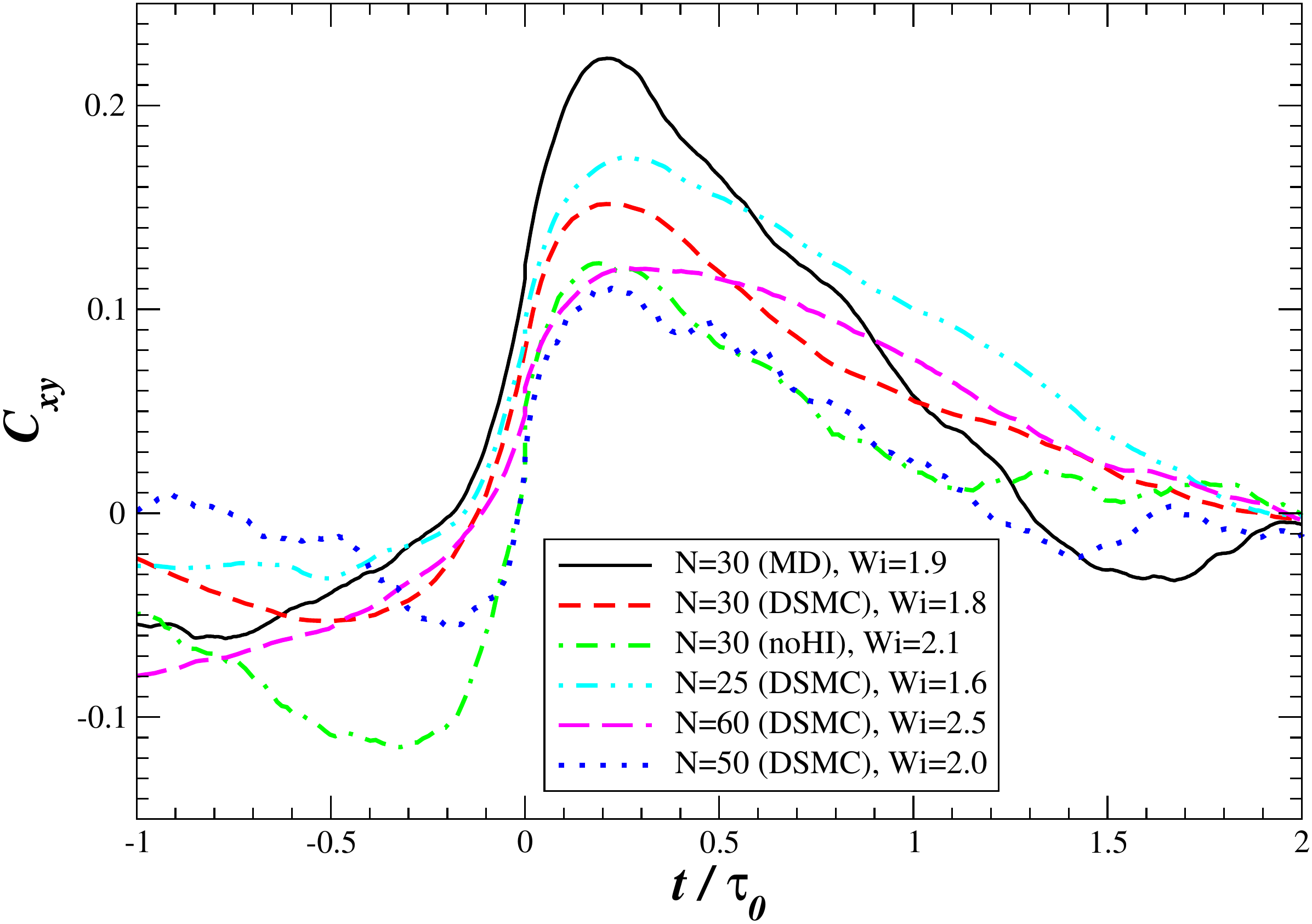}

\caption{\label{C_xy.Wi=3D2}Comparison of $C_{xy}(t)$ for Weissenberg number
of about $2$ for several different models, after the time axes has
been normalized. }
\end{figure}

In Fig. \ref{C_xy.Wi=3D2} we compare $C_{xy}(t)$ at $\Wi\approx2$
for several different models %
\footnote{The Weissenberg numbers were calculated after the runs were completed
and therefore the different runs are not at the exact same $\Wi$
number.%
} and see a good match, even for the DSMC runs ignoring hydrodynamics
(momentum conservation). This indicates that the dynamics of the chains
is primarily driven by the competition between the internal stochastic
motion (entropy) and the external forcing due to the shear, and not
hydrodynamic interactions between the beads or the effect of the motion
of the chain on the flow.

\begin{figure}
\includegraphics[width=0.75\columnwidth,keepaspectratio]{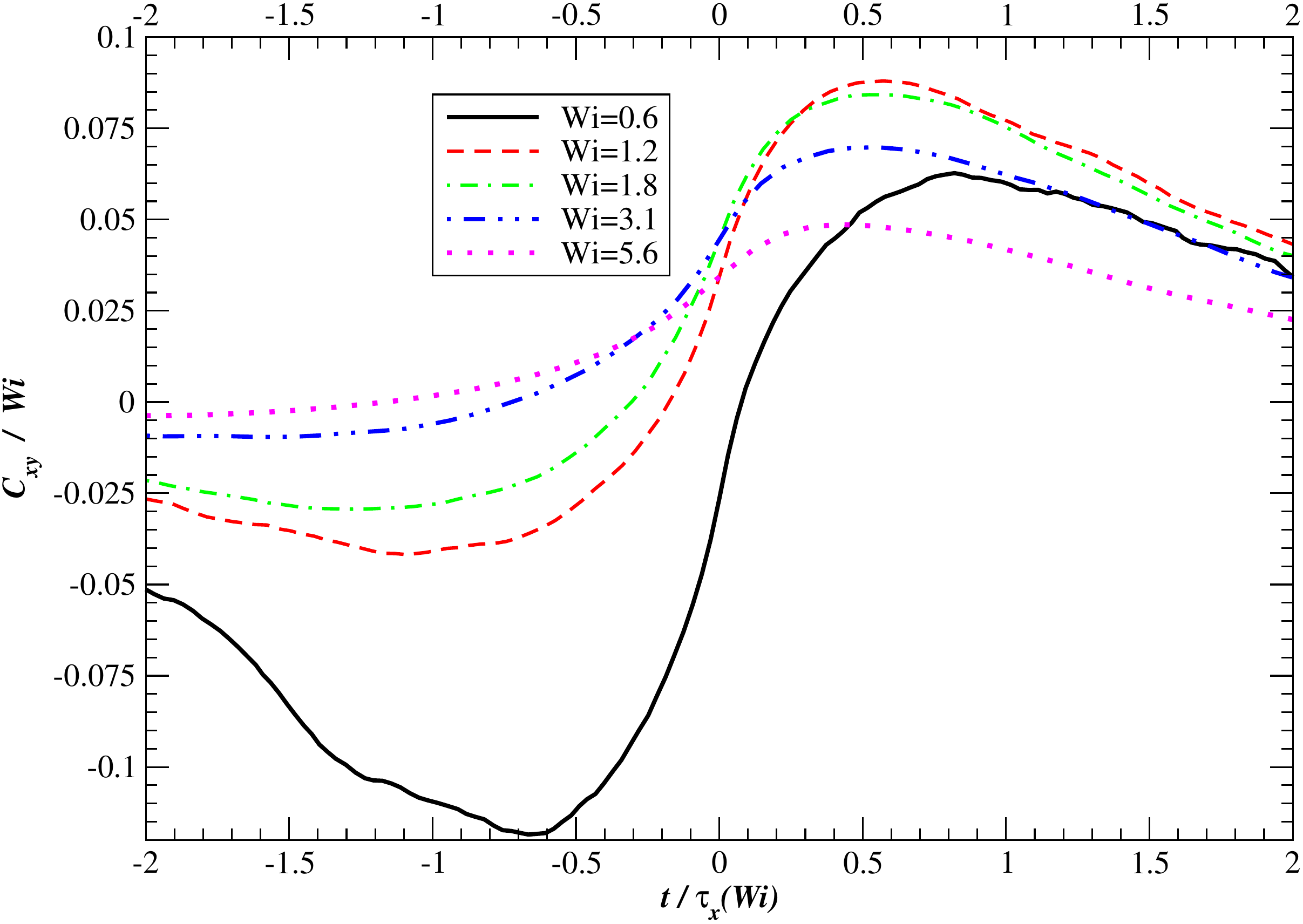}

\caption{\label{C_xy.renormalized}Cross correlation function $\tilde{C}_{xy}$
for $N=30$ DSMC runs as in Fig. \ref{C_xy.N=3D30} but with time
renormalized by $\tau(\Wi)$ and the correlation magnitude scaled
by $\Wi$. }
\end{figure}

We do not discuss the origin and locations of the peaks in the cross-correlation
functions in detail in this work. These peaks are indicative of the
existence of a correlated motion in the $xy$ plane, but do not uniquely
identify that motion. An important question to address is the existence
of a time scale other than the internal relaxation time $\tau(\Wi$).
In Fig. \ref{C_xy.renormalized} we show a renormalized cross-correlation
function \[
\tilde{C}_{xy}=\frac{1}{\Wi}C_{xy}\left[\frac{t}{\tau(\Wi)}\right]\]
in an unsuccessful attempt to collapse the data for different $\Wi$.
While the match is not perfect the picture does not point to the existence
of a time scale shorter than $\tau(\Wi)$. We also do not see any
convincing evidence for coherent and reproducible correlations on
time scales significantly larger than $\tau$, even in various power
spectral densities. Our results do not rule out the possibility of
a repetitive motion of the chain with widely varying cyclic times
(e.g., exponential tail) but we have not observed any direct evidence
for such cycling either. We will report more detailed results on the
dynamics of tethered polymer chains along with comparisons with Brownian
dynamics and Lattice-Boltzmann in future work.

\section{\label{sec:Conclusions}Conclusions}

We presented a stochastic event-driven molecular dynamics (SEDMD)
algorithm that combines hard-sphere event-driven molecular dynamics
(EDMD) with direct simulation Monte Carlo (DSMC), aimed at simulating
flow in suspensions at the microscale. The overall algorithm is still
event-driven, however, the DSMC portion of the algorithm can be made
time-driven for increased efficiency. The fundamental idea is to replace
the deterministic (MD-like) interactions between particles of certain
species with a stochastic (MC-like) collision process, thus preserving
the phase space dynamics and conservation laws but ignoring the liquid
structure. The SEDMD methodology correctly reproduces hydrodynamic
behavior at the macroscale but also correctly represents fluctuations
at the microscale. A similar algorithm has been proposed using time-driven
(soft-particle) MD and a multiparticle collision variant of DSMC \cite{DSMC_MPCD_MD_Kapral}.

As an application of such a methodology we have considered the simulation
of polymer chains in a flowing solution, and in particular, a polymer
tethered to a hard wall and subject to shear flow. We have implemented
open boundary conditions that adaptively adjust the simulation domain
to only focus on the region close to the polymer chain(s). The algorithm
is found to be efficient even though it is not parallelized, and it
is found to reproduce results obtained via molecular dynamics and
other algorithms in the literature, after adjusting for the correction
to transport coefficients and compressibility of the DSMC fluid relative
to the MD fluid.

We studied the dynamics of a tethered polymer subject to pure shear
and found consistent results between MD and DSMC and also previous
TDMD studies. We find that neither the size of the polymer beads relative
to the solvent particles, nor the correct representation of the hydrodynamic
interactions in the fluid, qualitatively alter the results. This suggests
that fluctuations dominate the dynamic behavior of tethered polymers,
consistent with previous studies. Our results do not find periodic
motion of the polymer and show that the cross-correlation between
the polymer extensions along the flow and shear directions shows a
double-peak structure with characteristic time that is comparable
to the relaxation time of the polymer. This is in contrast to other
works that claim the existence of a new timescale associated with
the cyclic motion of the polymer. We will investigate these issues
further and compare with Brownian dynamics and Lattice-Boltzmann simulations
in future work.

We expect that this and related algorithms will find many applications
in micro- and nano-fluidics. In particular, the use of DSMC instead
of expensive MD is suitable for problems where the detailed structure
and chemical specificity of the solvent do not matter, and more general
hydrodynamic forces and internal fluctuations dominate. Using a continuum
approach such as Navier-Stokes (NS) equations for the solvent is questionable
at very small length scales. Furthermore, the handling of singularities
and fluctuations is not natural in such PDE methods and various approximations
need to be evaluated using particle-based methods. Since the meshes
required by continuum solvers for microflows are very fine, it is
expected that the efficiency of particle methods will be comparable
to PDE solvers. Nevertheless, algorithms based on fluctuating hydrodynamics
descriptions will be more efficient when fluctuations matter. Comparisons
and coupling of DSMC to fluctuating NS solvers is the subject of current
investigations \cite{FluctuatingHydro_AMAR}.

\section{Acknowledgments}

This work was performed under the auspices of the U.S. Department
of Energy by the University of California Lawrence Livermore National
Laboratory under Contract No. W-7405-Eng-48 (UCRL-JRNL-233235).

\begin{appendix}

\section{\label{Appendix_AEDDSMC}AED variants of DSMC }

In this Appendix we discuss a fully asynchronous event-driven (AED)
implementation of DSMC. The advantage of asynchronous algorithms is
that they do not introduce any artificial time scales (such as a time
step) into the problem \cite{AED_Review}. We have validated that
the AED algorithm produces the same results as the time-driven one
by comparing against published DSMC results for plane Poiseuille flow
of a rare gas \cite{DSMC_PlanePoiseuille}. We have also implemented
traditional time-driven (TD) DSMC and find identical results when
the time step is sufficiently small. We find that the event-driven
algorithm is almost an order of magnitude slower than the time driven
one at higher densities, and only becomes competitive at very low
densities (which is the traditional domain of interest for DSMC).
The overhead of the AED algorithm comes from the need to re-predict
the next event and update the event queue whenever a particle suffers
a DSMC collision. This cost is in addition to the equivalent cost
in the time-driven algorithm, namely, moving the particles forward
in time and colliding them.

The AED algorithm introduces a new type of event, a \emph{stochastic
(trial) collision} between two DSMC particles that are in the same
cell (see Section \ref{Section_EDTD}). These trial collisions occur
in a given cell $c$ as a Poisson process (i.e., exponentially distributed
waiting time) with a rate given by Eq. (\ref{eq:Gamma_c}). There
are several approaches to scheduling and processing DSMC collisions
directly borrowed from algorithms for performing Kinetic Monte Carlo
simulations (which are \emph{synchronous} event-driven algorithms
\cite{AdvancedKMC_Tutorial}). The simplest, and in our experience,
most efficient, approach to AED DSMC is to use \emph{cell rejection}
to select a host cell for the stochastic collisions. The rate of DSMC
collisions is chosen according to the cell with maximal occupancy
$N_{c}^{max}$, $\Gamma=N_{cells}\Gamma_{c}^{max}$. The randomly
chosen cell $c$ of occupancy $N_{c}$ is accepted with probability
$N_{c}(N_{c}-1)/\left[N_{c}^{max}(N_{c}^{max}-1)\right]$ and a random
pair of particles $i$ and $j$ are chosen from $\mathcal{L}_{c}$.
Since the DSMC fluid is perfectly compressible, the maximal cell occupancy
can be quite high for very large systems, and this leads to decreasing
cell acceptance probability as the size of the system increases.

One can avoid cell rejections altogether. The first option is to associate
stochastic collisions with cells and schedule one such event per cell.
The event time is easily predicted at any point in time $t$ to occur
at time $t-\Gamma_{c}^{-1}\ln r$ , where $r$ is a uniform random
deviate in $(0,1)$. These event times are put in the event queue
(which may be separate from the particle one and then the two queues
may be merged only at the top). The event times (and thus the cell
queue) need to updated whenever a cell occupancy $N_{c}$ changes,
that is, whenever a cell transfer is processed. This makes this algorithm
inefficient. Another alternative is to recognize that the sum of a
set of independent Poisson processes is a Poisson process with a rate
that is the sum of the individual rates, $\Gamma=\sum_{c}\Gamma_{c}$.
That is, DSMC collisions occur in the system as a Poisson process
with rate $\Gamma$. When processing such an event one has to first
choose the cell with probability $\Gamma_{c}/\Gamma$, which requires
some additional data structures to implement efficiently \cite{AdvancedKMC_Tutorial}.
For example, the cells could be grouped in lists based on their occupancy
and then an occupancy chosen first (with the appropriate weight),
followed by selection of a cell with that particular occupancy.

Finally, it is also possible to use a mixture of the asynchronous
and time-driven variants of DSMC. The asynchronous algorithm (e.g.,
based on cell rejection) can be used for DSMC particles in event-driven
cells, and the time-driven one elsewhere. This may be useful in situations
where the time-scale of the event-driven component (i.e., the solute
and nearby solvent particles) is significantly smaller than the time
step $\D{t}$ and thus time stepping would lead to discretization
artifacts. 

In the AED variant of DSMC constant pressure BCs (see Section \ref{Section_ConstantPressure})
can be implemented by adding a new type of \emph{acceleration event}.
When such an event is processed, all of the particles are brought
to the same point in time (synchronized), the velocities of each DSMC
particle $i$ is incremented by $\V{a}\D{t}_{i}$ (here $\D{t}_{i}$
is the elapsed time since the last acceleration event), and the event
handling is restarted. The acceleration events occur as a Poisson
process with a suitably chosen rate, for example, ensuring that the
average or maximal change in velocity is a fraction of the average
particle velocity. Note that the choice of this acceleration rate
introduces an artificial time constant in the algorithm similar to
the time step $\D{t}$ in time-driven DSMC.

\section{\label{Appendix_TRMC}Multi-Particle Collisions in DSMC}

Under dense liquid conditions, DSMC binary collisions are so numerous
(see Section \ref{Section_Gamma_c}) that the velocities of the particles
are effectively thermalized to the local Maxwell distribution. We
have implemented a variant DSMC algorithm in which at every time step
the velocities of all of the particles are redrawn from a local Maxwellian,
preserving the total linear momentum and energy in each cell \cite{DSMC_Pullin}.
We found that this variant of DSMC is less efficient than and behaves
similarly to the usual binary-collision DSMC. Reference \cite{DSMC_TRMC}
describes a more general algorithm (TRMC) that combines binary collisions
for a subset of the particles with drawing from a local Maxwellian
for the remainder of the particles, and under dense liquid conditions
this typically degenerates to complete randomization of all of the
velocities at every time step. Until a theoretical framework is established
for the behavior of DSMC-like algorithms at high densities the classical
DSMC algorithm seems to be the best alternative in terms of simplicity,
efficiency, and theoretical foundation. The effect of collision rules
and cell size on multiparticle collision dynamics has been studied
and it was found that increased collisional viscosity is desirable
for achieving realistic convection to diffusion ratios \cite{DSMC_MPCD_Gompper,MultiparticleDSMC_Polymers}.

We mention that, strictly speaking, we should use as $V_{c}$ in Eq.
(\ref{eq:Gamma_c}) not the volume of the cell, but the unoccupied
cell volume (that is, the portion of the cell not covered by non-DSMC
particles) %
\footnote{Our implementation makes the additional approximation that non-DSMC
particles are also counted in $N_{c}$ in Eq. (\ref{eq:Gamma_c}),
instead of keeping a separate count of just the DSMC particles inside
each cell. If the polymer beads are larger than a cell than this approximation
does not matter since no cell can contain the centroid of both a DSMC
and non-DSMC particle.%
}. It is however difficult to dynamically maintain an accurate estimate
of the cell coverage, and the complication does not appear to be worth
the implementation complexity. In particular, an approximation is
already made in neglecting the structure of the fluid near a polymer
bead (i.e., the solvation layer), and furthermore, the majority of
the cells that are partially covered by a polymer bead will be entirely
or almost entirely covered so that they would at most contain a single
DSMC particle, in which case the probability of a DSMC collision would
be very low anyway. Finally, as explained in Section \ref{Section_Gamma_c},
the exact collision frequency does not really matter. In the context
of multiparticle collision dynamics, Ref. \cite{DSMC_MPCD_CylinderFlow}
proposes the use of virtual particles filling the partially-filled
cells as a way to achieve more accurate stick boundary conditions.

\end{appendix}

\bibliographystyle{unsrt}
\bibliography{6_home_donev1_HPC_Papers_DSMC_References}

\begin{thebibliography}{10}

\bibitem{Nanohydrodynamics_Alder}
W.~E. Alley, P.~Covello, and B.~J. Alder.
\newblock {Complex flows by nanohydrodynamics}.
\newblock {\em Mol. Phys.}, 102(19):0026--8976, 2004.

\bibitem{DSMC_MPCD_Gompper}
M.~Ripoll, K.~Mussawisade, R.~G. Winkler, and G.~Gompper.
\newblock {Low-Reynolds-number hydrodynamics of complex fluids by
  multi-particle-collision dynamics}.
\newblock {\em Europhys. Lett.}, 68(1):106, 2004.

\bibitem{Microfluidics_Review}
T.~M. Squires and S.~R. Quake.
\newblock {Microfluidics: Fluid physics at the nanoliter scale}.
\newblock {\em Rev. Mod. Phys.}, 77(3):977, 2005.

\bibitem{PolymerTumbling_Review}
E.~S.~G. Shaqfeh.
\newblock {The dynamics of single-molecule DNA in flow}.
\newblock {\em Journal of Non-Newtonian Fluid Mechanics}, 130:1--28, 2005.

\bibitem{FluctuatingHydro_Coveney}
G.~De Fabritiis, M.~Serrano, R.~Delgado-Buscalioni, and P.~V. Coveney.
\newblock {Fluctuating hydrodynamic modeling of fluids at the nanoscale}.
\newblock {\em Phys. Rev. E}, 75(2):026307, 2007.

\bibitem{PolymerDynamics_Review}
G.~W. Slater, Y.~Gratton, M.~Kenward, L.~McCormick, and F.~Tessier.
\newblock {Deformation, Stretching, and Relaxation of Single-Polymer Chains:
  Fundamentals and Examples}.
\newblock {\em Soft Materials}, 2:155--182, 2004.

\bibitem{TetheredPolymer_Experiment_PRL}
P.~S. Doyle, B.~Ladoux, and J.-L. Viovy.
\newblock {Dynamics of a Tethered Polymer in Shear Flow}.
\newblock {\em Phys. Rev. Lett.}, 84(20):4769--4772, 2000.

\bibitem{TetheredPolymer_HybridMD}
S.~Barsky, R.~Delgado-Buscalioni, and P.~V. Coveney.
\newblock {Comparison of molecular dynamics with hybrid continuum--molecular
  dynamics for a single tethered polymer in a solvent}.
\newblock {\em J. Chem. Phys.}, 121(5):2403--2411, 2004.

\bibitem{TetheredPolymer_FullMD}
Y.~Gratton and G.~W. Slater.
\newblock {Molecular dynamics study of tethered polymers in shear flow}.
\newblock {\em Eur. Phys. J. E}, 17:455--465, 2005.

\bibitem{TetheredPolymer_Cyclic_PRL}
R.~Delgado-Buscalioni.
\newblock {Cyclic Motion of a Grafted Polymer under Shear Flow}.
\newblock {\em Phys. Rev. Lett.}, 96(8):088303, 2006.

\bibitem{TetheredDNA_FullMD}
G.~M Wang and W.~C Sandberg.
\newblock {Non-equilibrium all-atom molecular dynamics simulations of free and
  tethered DNA molecules in nanochannel shear flows}.
\newblock {\em Nanotechnology}, 18(13):135702, 2007.

\bibitem{PolymerTumbling_PRL}
C.~M. Schroeder, R.~E. Teixeira, E.~S.~G. Shaqfeh, and S.~Chu.
\newblock {Characteristic Periodic Motion of Polymers in Shear Flow}.
\newblock {\em Phys. Rev. Lett.}, 95(1):018301, 2005.

\bibitem{TetheredPolymer_Cyclic_AIP}
R.~Delgado-Buscalioni.
\newblock {Dynamics of a Single Tethered Polymer under Shear Flow}.
\newblock {\em AIP Conference Proceedings}, 913(1):114--120, 2007.

\bibitem{BrownianDynamics_DNA}
R.~M. Jendrejack, J.~J. de~Pablo, and M.~D. Graham.
\newblock {Stochastic simulations of DNA in flow: Dynamics and the effects of
  hydrodynamic interactions}.
\newblock {\em J. Chem. Phys.}, 116(17):7752--7759, 2002.

\bibitem{BrownianDynamics_OrderN}
J.~P. Hernandez-Ortiz, J.~J. de~Pablo, and M.~D. Graham.
\newblock {Fast Computation of Many-Particle Hydrodynamic and Electrostatic
  Interactions in a Confined Geometry}.
\newblock {\em Phys. Rev. Lett.}, 98(14):140602, 2007.

\bibitem{LatticeBoltzmann_Polymers}
O.~B. Usta, A.~J.~C. Ladd, and J.~E. Butler.
\newblock {Lattice-Boltzmann simulations of the dynamics of polymer solutions
  in periodic and confined geometries}.
\newblock {\em J. Chem. Phys.}, 122(9):094902, 2005.

\bibitem{FluctuatingHydro_FluidOnly}
N.~Sharma and N.~A. Patankar.
\newblock {Direct numerical simulation of the Brownian motion of particles by
  using fluctuating hydrodynamic equations}.
\newblock {\em J. Comput. Phys.}, 201:466--486, 2004.

\bibitem{DNA_Laden_Flow}
D.~Trebotich, G.~H. Miller, P.~Colella, D.~T. Graves, D.~F. Martin, and P.~O.
  Schwartz.
\newblock {A Tightly Coupled Particle-Fluid Model for DNA-Laden Flows in
  Complex Microscale Geometries}.
\newblock {\em Comp. Fluid Solid Mech.}, pages 1018--1022, 2005.

\bibitem{FluctuatingHydroMD_Coveney}
G.~Giupponi, G.~De Fabritiis, and P.~V. Coveney.
\newblock {Hybrid method coupling fluctuating hydrodynamics and molecular
  dynamics for the simulation of macromolecules}.
\newblock {\em J. Chem. Phys.}, 126(15):154903, 2007.

\bibitem{PolymerCollapse_Yeomans}
N.~Kikuchi, J.~F. Ryder, C.~M. Pooley, and J.~M. Yeomans.
\newblock {Kinetics of the polymer collapse transition: The role of
  hydrodynamics}.
\newblock {\em Phys. Rev. E}, 71(6):061804, 2005.

\bibitem{Trebotich_HardRods}
D.~Trebotich, G.~H. Miller, and M.~D. Bybee.
\newblock {A Hard Constraint Algorithm to Model Particle Interactions in
  DNA-laden Flows}.
\newblock {\em Nanoscale and Microscale Thermophysical Engineering},
  11(1):121--128, 2007.

\bibitem{MultiparticleDSMC_Polymers}
K.~Mussawisade, M.~Ripoll, R.~G. Winkler, and G.~Gompper.
\newblock {Dynamics of polymers in a particle-based mesoscopic solvent}.
\newblock {\em J. Chem. Phys.}, 123(14):144905, 2005.

\bibitem{DSMC_MPCD_MD_Kapral}
S.~H. Lee and R.~Kapral.
\newblock {Mesoscopic description of solvent effects on polymer dynamics}.
\newblock {\em J. Chem. Phys.}, 124(21):214901, 2006.

\bibitem{Trebotich_Penalty}
D.~Trebotich, G.~H. Miller, and M.~D. Bybee.
\newblock {A Penalty Method to Model Particle Interactions in DNA-laden Flows}.
\newblock To appear in J. Nanosci. Nanotech., 2007.

\bibitem{PolymerShear_MD}
C.~Aust, M.~Kroger, and S.~Hess.
\newblock Structure and dynamics of dilute polymer solutions under shear flow
  via nonequilibrium molecular dynamics.
\newblock {\em Macromolecules}, 32(17):5660--5672, 1999.

\bibitem{FluctuatingHydro_Garcia}
J.~B. Bell, A.~Garcia, and S.~A. Williams.
\newblock {Numerical Methods for the Stochastic Landau-Lifshitz Navier-Stokes
  Equations}.
\newblock {\em Phys. Rev. E}, 76:016708, 2007.

\bibitem{FluctuatingHydro_AMAR}
S.~A. Williams, J.~B. Bell, and A.~L. Garcia.
\newblock {Algorithm Refinement for Fluctuating Hydrodynamics}.
\newblock Submitted, 2007.

\bibitem{FluidMixing_DSMC}
K.~Kadau, C.~Rosenblatt, J.~L. Barber, T.~C. Germann, Z.~Huang, P.~Carles, and
  B.~J. Alder.
\newblock {The importance of fluctuations in fluid mixing}.
\newblock {\em PNAS}, 104(19):7741--7745, 2007.

\bibitem{DPD_DNA}
F.~Xijunand~N. Phan-Thien, S.~Chen, X.~Wu, and T.~Y. Ng.
\newblock {Simulating flow of DNA suspension using dissipative particle
  dynamics}.
\newblock {\em Physics of Fluids}, 18(6):063102, 2006.

\bibitem{DSMCReview_Garcia}
F.~J. Alexander and A.~L. Garcia.
\newblock {The Direct Simulation Monte Carlo Method}.
\newblock {\em Computers in Physics}, 11(6):588--593, 1997.

\bibitem{DSMC_DenseFluids}
F.~Baras, M.~Malek Mansour, and A.~L. Garcia.
\newblock {Microscopic simulation of dilute gases with adjustable transport
  coefficients}.
\newblock {\em Phys. Rev. E}, 49(4):3512--3515, 1994.

\bibitem{DSMC_CBA}
F.~J. Alexander, A.~L. Garcia, and B.~J. Alder.
\newblock {A Consistent Boltzmann Algorithm}.
\newblock {\em Phys. Rev. Lett.}, 74(26):5212--5215, 1995.

\bibitem{DSMC_CBATheory}
A.~L. Garcia and W.~Wagner.
\newblock {The limiting kinetic equation of the Consistent Boltzmann Algorithm
  for dense gases}.
\newblock {\em J. Stat. Phys.}, 101:1065--86, 2000.

\bibitem{ShearThinning_Yeomans}
J.~F. Ryder and J.~M. Yeomans.
\newblock Shear thinning in dilute polymer solutions.
\newblock {\em The Journal of Chemical Physics}, 125(19):194906, 2006.

\bibitem{DSMC_Enskog}
Jose~Maria Montanero and Andres Santos.
\newblock {Simulation of the Enskog equation [a-grave] la Bird}.
\newblock {\em Phys. Fluids}, 9(7):2057--2060, 1997.

\bibitem{DSMC_Enskog_Frezzotti}
A.~Frezzotti.
\newblock {A particle scheme for the numerical solution of the Enskog
  equation}.
\newblock {\em Phys. Fluids}, 9(5):1329--1335, 1997.

\bibitem{MPCD_CBA}
T.~Ihle, E.~Tüzel, and D.~M. Kroll.
\newblock {Consistent particle-based algorithm with a non-ideal equation of
  state}.
\newblock {\em Europhys. Lett.}, 73:664--670, 2006.

\bibitem{EventDriven_Alder}
B.~J. Alder and T.~E. Wainwright.
\newblock {Studies in molecular dynamics. I. General method}.
\newblock {\em J. Chem. Phys.}, 31:459, 1959.

\bibitem{Event_Driven_HE}
A.~Donev, S.~Torquato, and F.~H. Stillinger.
\newblock {Neighbor List Collision-Driven Molecular Dynamics Simulation for
  Nonspherical Particles: {I.} Algorithmic Details {II.} Applications to
  Ellipses and Ellipsoids}.
\newblock {\em J. Comp. Phys.}, 202(2):737--764, 765--793, 2005.

\bibitem{EDMD_Polymers_Hall}
S.~W. Smith, C.~K. Hall, and B.~D. Freeman.
\newblock {Molecular Dynamics for Polymeric Fluids Using Discontinuous
  Potentials}.
\newblock {\em J. Comp. Phys.}, 134(1):16--30, 1997.

\bibitem{PolymerCollapse_EDMD}
S.~B. Opps, J.~M. Polson, and N.~A. Risk.
\newblock {Discontinuous molecular dynamics simulation study of polymer
  collapse}.
\newblock {\em J. Chem. Phys.}, 125(19):194904, 2006.

\bibitem{EDMD_Polymers_Aggregation}
S.~Peng, F.~Ding, B.~Urbanc, S.~V. Buldyrev, L.~Cruz, H.~E. Stanley, and N.~V.
  Dokholyan.
\newblock Discrete molecular dynamics simulations of peptide aggregation.
\newblock {\em Phys. Rev. E}, 69(4):041908, 2004.

\bibitem{EDMD_Polymer_Fibrils2}
H.~D. Nguyen and C.~K. Hall.
\newblock {Molecular Dynamics Simulations of Fibril Formation by Random Coil
  Peptides}.
\newblock {\em Proc. Natl. Acad. Sci.}, USA 101:16180, 2004.

\bibitem{AED_Serial}
A.~Donev.
\newblock {Asynchronous Event-Driven Particle Algorithms}.
\newblock In K.~S. Perumalla, editor, {\em {Proceedings of the 21st
  International Workshop on Principles of Advanced and Distributed
  Simulation}}. IEEE Computer Society, June 2007.

\bibitem{DSMC_TimeStepError}
A.~L. Garcia and W.~Wagner.
\newblock {Time step truncation error in direct simulation Monte Carlo}.
\newblock {\em Phys. Fluids}, 12:2621--2633, 2000.

\bibitem{DSMC_InflowDistribution}
A.~L. Garcia and W.~Wagner.
\newblock {Generation of the Maxwellian Inflow Distribution}.
\newblock {\em J. Comput. Phys.}, 217:693--708, 2006.

\bibitem{AMAR_DSMC}
A.~L. Garcia, J.~Bell, Wm.~Y. Crutchfield, and B.~J. Alder.
\newblock {Adaptive Mesh and Algorithm Refinement using Direct Simulation Monte
  Carlo}.
\newblock {\em J. Comp. Phys.}, 154:134--155, 1999.

\bibitem{AMAR_DSMC_SAMRAI}
S.~Wijesinghe, R.~Hornung, A.~L. Garcia, and N.~Hadjiconstantinou.
\newblock {Three-dimensional Hybrid Continuum-Atomistic Simulations for
  Multiscale Hydrodynamics}.
\newblock {\em Journal of Fluids Engineering}, 126:768--777, 2004.

\bibitem{DSMC_MPCD_CylinderFlow}
A.~Lamura, G.~Gompper, T.~Ihle, and D.~M. Kroll.
\newblock {Multi-particle collision dynamics: Flow around a circular and a
  square cylinder}.
\newblock {\em Europhys. Lett.}, 56(3):319--325, 2001.

\bibitem{SuspensionsDSMC_Reflection}
Y.~Inoue, Y.~Chen, and H.~Ohashi.
\newblock {Development of a Simulation Model for Solid Objects Suspended in a
  Fluctuating Fluid}.
\newblock {\em J. Stat. Phys.}, 107(112):85--100, 2002.

\bibitem{DSMC_PlanePoiseuille}
F.~J. Uribe and A.~L. Garcia.
\newblock Burnett description for plane poiseuille flow.
\newblock {\em Phys. Rev. E}, 60(4):4063--4078, 1999.

\bibitem{DSMC_CellSizeError}
F.~Alexander, A.~L. Garcia, and B.~J. Alder.
\newblock {Cell Size Dependence of Transport Coefficients in Stochastic
  Particle Algorithms}.
\newblock {\em Phys. Fluids}, 10:1540, 1998.
\newblock Erratum: Phys. Fluids, 12 731 (2000).

\bibitem{DSMC_GammaBounds}
T.~Bartel, T.~Sterk, J.~Payne, and B.~Preppernau.
\newblock {DSMC simulation of nozzle expansion flow fields}.
\newblock AIAA Paper No. 94-2047, 1994.

\bibitem{PolymerTumbling_PDFs}
S.~Gerashchenko and V.~Steinberg.
\newblock {Statistics of Tumbling of a Single Polymer Molecule in Shear Flow}.
\newblock {\em Phys. Rev. Lett.}, 96(3):038304, 2006.

\bibitem{AED_Review}
Aleksandar Donev.
\newblock Asynchronous event-driven particle algorithms.
\newblock In {\em PADS '07: Proceedings of the 21st International Workshop on
  Principles of Advanced and Distributed Simulation}, pages 83--92, Washington,
  DC, USA, 2007. IEEE Computer Society.

\bibitem{AdvancedKMC_Tutorial}
M.~A. Novotny.
\newblock {\em {A Tutorial on Advanced Dynamic Monte Carlo Methods for Systems
  with Discrete State Spaces}}, volume~IX of {\em Annual Reviews of
  Computational Physics}, pages 153--210.
\newblock World Scientific, Singapore, 2001.

\bibitem{DSMC_Pullin}
D.~I. Pullin.
\newblock {Direct simulation methods for compressible inviscid ideal-gas flow}.
\newblock {\em J. Comp. Phys.}, 34:231--244, 1980.

\bibitem{DSMC_TRMC}
L.~Pareschi and S.Trazzi.
\newblock {Numerical solution of the Boltzmann equation by Time Relaxed Monte
  Carlo (TRMC) methods}.
\newblock {\em Int. J. Numer. Meth. Fluids}, 48:947--983, 2005.

\end{thebibliography}

\end{document}